\documentclass[12pt, reqno]{amsart}

\usepackage[cp1251]{inputenc}

\usepackage[russian, english]{babel}

\usepackage[margin=1.in, a4paper, foot=.3in]{geometry}

\usepackage{graphicx, color}

\definecolor{darkred}{rgb}{.5,0,0}
\definecolor{darkgreen}{rgb}{0,0.3,0}
\definecolor{darkblue}{rgb}{0,0,.5}

\usepackage{hyperref}
\hypersetup {
    pdfstartview=FitH,
    colorlinks,
    citecolor=darkgreen,
    linkcolor=darkred,
    urlcolor=darkblue
}

\usepackage{cite}

\usepackage[noBBpl,slantedGreek]{mathpazo}
\usepackage[mathcal]{euscript}

\usepackage{bm}

\usepackage{amssymb}
\usepackage{amsmath}
\usepackage{stmaryrd}
\usepackage{mathtools}

\allowdisplaybreaks

\numberwithin{equation}{section}

\newcommand {\ms}{{\mathstrut}} 

\newcommand {\bbC}{\mathbb C}
\newcommand {\bbN}{\mathbb N}
\newcommand {\bbZ}{\mathbb Z}

\newcommand {\calO}{\mathcal O}

\newcommand {\gothh}{\mathfrak h}

\newcommand {\hgothh}{\widehat{\mathfrak h}}
\newcommand {\tgothh}{\widetilde{\mathfrak h}}

\newcommand {\hi}{\widehat I}

\newcommand {\gllpo}{\mathfrak{gl}_{l+1}}
\newcommand {\sllpo}{\mathfrak{sl}_{l+1}}
\newcommand {\lsllpo}{{\mathcal L}({\mathfrak{sl}}_{l+1})}
\newcommand {\tlsllpo}{\widetilde{\mathcal L}({\mathfrak{sl}}_{l+1})}
\newcommand {\hlsllpo}{\widehat{\mathcal L}(\mathfrak{sl}_{l + 1})}
\newcommand {\uhlsllpo}{\mathrm U(\widehat{\mathcal L}(\mathfrak{sl}_{l + 1}))}

\newcommand {\uqlsllpo}{\mathrm U_q(\mathcal L(\mathfrak{sl}_{l + 1}))}
\newcommand {\uqtlsllpo}{\mathrm U_q(\widetilde{\mathcal L}(\mathfrak{sl}_{l + 1}))}
\newcommand {\uqhlsllpo}{\mathrm U_q(\widehat{\mathcal L}(\mathfrak{sl}_{l + 1}))}

\newcommand {\uqbm}{\mathrm U_q(\mathfrak b_-)}
\newcommand {\uqbp}{\mathrm U_q(\mathfrak b_+)}

\newcommand {\uqgllpo}{\mathrm U_q(\mathfrak{gl}_{l + 1})}

\newcommand {\mbar}[3]{\hskip #2 \overline{\hskip -#2 #1 \hskip -#3} \hskip #3}

\newcommand {\ovlambda}{\mbar{\lambda}{.03em}{.03em}}
\newcommand {\ovotimes}{\mathbin{\mbar{\otimes}{.1em}{.1em}}}
\newcommand {\ovrho}{\mbar{\rho}{.03em}{.03em}}
\newcommand {\ovPsi}{\mbar{\Psi}{.03em}{.03em}}
\newcommand {\ovtheta}{\mbar{\theta}{.03em}{.03em}}
\newcommand {\ovv}{\mbar{v}{.05em}{.01em}}

\newcommand {\ovQ}{\mbar{Q}{.03em}{.03em}}

\DeclareMathOperator {\Osc}{Osc}
\DeclareMathOperator {\id}{\mathrm{id}}

\title[Oscillator versus prefundamental representations II. Arbitrary higher ranks]{Oscillator versus prefundamental representations II. \\ Arbitrary higher ranks}

\author[H. Boos]{Hermann Boos}
\address{Mathematics and Natural Sciences, University of Wuppertal, 42097 Wuppertal, Germany}
\email{hboos@uni-wuppertal.de}

\author[F. G\"ohmann]{Frank G\"ohmann}
\address{Mathematics and Natural Sciences, University of Wuppertal, 42097 Wuppertal, Germany}
\email{goehmann@uni-wuppertal.de}

\author[A. Kl\"umper]{Andreas Kl\"umper}
\address{Mathematics and Natural Sciences, University of Wuppertal, 42097 Wuppertal, Germany}
\email{kluemper@uni-wuppertal.de}

\author[Kh. S. Nirov]{\vskip .2em Khazret S. Nirov}
\address{Institute for Nuclear Research of the Russian Academy of Sciences, 60th October Ave 7a, 117312 Moscow, Russia}
\curraddr{Mathematics and Natural Sciences, University of Wuppertal, 42097 Wuppertal, Germany}
\email{nirov@uni-wuppertal.de \& nirov@inr.ac.ru}

\author[A. V. Razumov]{Alexander V. Razumov}
\address{Institute for High Energy Physics, NRC "Kurchatov Institute", 142281 Protvino, Mos\-cow region, Russia}
\email{Alexander.Razumov@ihep.ru}

\begin{document}

\addtolength {\jot}{3pt}

\begin{abstract}
We find the $\ell$-weights and the $\ell$-weight vectors for the highest $\ell$-weight  $q$-oscillator representations of the positive Borel subalgebra of the quantum loop algebra $U_q(\mathcal L(\mathfrak{sl}_{l+1}))$ for arbitrary values of $l$. Having this, we establish the explicit relationship between the $q$-oscillator and prefundamental representations. Our consideration allows us to conclude that the prefundamental representations can be obtained by tensoring $q$-oscillator representations.
\end{abstract}

\maketitle

\tableofcontents

\section{Introduction}

Within the framework of the group-theoretic approach to quantum integrable systems the role of the symmetry groups and algebras is played by the quantum group. The corresponding notion was formulated by Drinfeld \cite{Dri85, Dri87} and 
Jimbo \cite{Jim85}, both motivated by the quantum inverse scattering method. 

The modern development of the quantum group approach was mainly due to a 
series of remarkable papers \cite{BazLukZam96, BazLukZam97, BazLukZam99} 
and \cite{BazHibKho02}, where the quantum versions of the KdV equation and the 
simplest boundary affine Toda field theory, respectively, were considered. The 
key object here is the universal $R$-matrix formally introduced as an element 
of the tensor product of two copies of the quantum group. It has become 
traditional to call the representation used for the first factor of this tensor product the auxiliary space, 
while treating the representation space of the second one as the quantum space. These conventions can also be 
interchanged. Important feature of the method is that, by choosing a representation of the quantum group in the 
auxiliary space, one fixes an integrability object being either a monodromy- or a transfer-type operator, 
while subsequently fixing a representation in the quantum space one defines a physical model, such as a 
low-dimensional quantum field theory \cite{BazLukZam96, BazLukZam97, BazLukZam99, BazHibKho02} or a spin-chain 
model \cite{BooGoeKluNirRaz14a, BooGoeKluNirRaz14b, NirRaz16a}. And moreover, the functional relations between 
integrability objects should also follow from the characteristics of these 
representations of the quantum group. In algebraic aspects, the investigation 
of a quantum integrable system is thus reduced to the study of representations
of the quantum group to which the system is associated. For terminology, specific 
examples and references we refer here to the paper \cite{BooGoeKluNirRaz14a}.

For further constructions, it is essential to clarify that the universal $R$-matrix 
belongs not to the tensor square of the whole quantum group, but to a completed tensor 
product of its principal Borel subalgebras. Therefore, one can start with a representation 
of the quantum group and obtain desirable representations of its positive and negative 
Borel subalgebras by the corresponding restrictions. Actually, this is an obvious way 
to construct monodromy and transfer operators. To obtain more representations, leading 
to other integrability objects, such as $L$-operators and $Q$-operators, one should follow 
a different way. One possibility here is to map the Borel subalgebra in question to 
a $q$-oscillator algebra subsequently using the representations of the latter. 
Representations obtained in this way are connected with the representations used 
for the construction of monodromy and transfer operators through a certain limiting 
procedure \cite{BazHibKho02, BooGoeKluNirRaz13, BooGoeKluNirRaz14a, BooGoeKluNirRaz14b, NirRaz16a}. 
This fact ensures nontrivial functional relations between integrability objects.

Special attention in our research program is payed to quantum groups of loop algebras
of special linear Lie algebras, the so-called quantum loop algebras, and to their Borel
subalgebras. For the study of their representations the concept of $\ell$-weights and 
$\ell$-weight vectors turns out to be very productive \cite{HerJim12, FreHer15, MukYou14}. 
Earlier, specific $\ell$-weights emerged in \cite{ChaPre91, ChaPre94} in the 
classification of irreducible finite-dimensional representations of quantum loop 
algebras; in \cite{ChaPre94}, they were called pseudo-highest weights. Quite recently, 
some new infinite-dimensional representations of the Borel subalgebras of quantum loop 
algebras were obtained as an infinite limit of the Kirillov--Reshetikhin modules \cite{HerJim12}. 
Following \cite{FreHer15}, where relations in the Grothendieck ring of the category $\calO$ were 
interpreted as generalized Baxter's $TQ$-relations, we call these representations prefundamental. 
As the Kirillov--Reshetikhin modules are finite-dimensional highest weight modules characterized 
by specially simple values of the weights, so the infinite-dimensional prefundamental representations 
are highest $\ell$-weight modules characterized by their highest $\ell$-weights which are actually 
of simplest possible form. 

In our recent paper \cite{BooGoeKluNirRaz16}, we obtained the $\ell$-weights and 
the corresponding $\ell$-weight vectors for the finite- and infinite-dimensional 
representations of quantum loop algebras $\uqlsllpo$ with $l = 1, 2$ constructed 
with the help of the evaluation representations. There, we also obtained the 
$\ell$-weights and the $\ell$-weight vectors for the $q$-oscillator representations 
of the positive Borel subalgebras of the same quantum loop algebras. In particular, 
this allowed us to relate the $q$-oscillator and prefundamental representations. The 
consideration of the lowest, $l = 1$, and the simplest higher rank, $l = 2$, cases 
in the paper \cite{BooGoeKluNirRaz16} does not, however, allow us to make even an 
educated guess on the exact relationship between the $q$-oscillator and prefundamental 
representations for arbitrary higher ranks. Therefore, to establish the complete 
relationship between these two important representations of quantum loop algebras, 
we must consider the general case, the quantum loop algebra  $\uqlsllpo$ for arbitrary 
$l$. To this end, we can use the results of the paper \cite{NirRaz16b}, where Borel 
subalgebras of these quantum loop algebras were realized by means of $q$-oscillators.

In section 2 we recall two different approaches to the definition of the quantum
loop algebras, one according to original definition of quantum groups by Drinfeld 
and Jimbo, and another one being Drinfeld's second realization of the same
object. We also provide the bijection between these two definitions explicitly
relating the generators of Drinfeld's second realization with the Cartan--Weyl
generators of Drinfeld--Jimbo's definition. In section 3 we define highest 
$\ell$-weight representations of Borel subalgebras of the quantum loop algebras. 
After a short reminder of general data on such representations, we introduce two 
corresponding examples of our particular interest, the prefundamental and 
$q$-oscillator representations. In section 4 we extend the set of $q$-oscillator
representations with the help of automorphisms of the Borel subalgebras under
consideration. Subsequently, for each $q$-oscillator representation we obtain 
the $\ell$-weight vectors and calculate the corresponding $\ell$-weights. These
results allow us to compare the $q$-oscillator representations directly with the
prefundamental representations. Our conclusions are gathered in the final section. 

We follow the definitions used in \cite{BooGoeKluNirRaz16}, and this is what is 
usually adopted in the papers on representations of quantum loop algebras. The 
so-called deformation parameter is introduced as follows. We take $\hbar$ to be 
a nonzero complex number, such that $q = \exp \hbar$ is not a root of unity. In 
this way, the quantum groups under consideration are defined simply as complex 
algebras. It is also assumed here that 
\[
q^\nu = \exp (\hbar \nu)
\]
for any $\nu \in \bbC$. We use the usual definition of $q$-number,
\begin{equation*}
[n]_q = \frac{q^n - q^{-n}}{q - q^{-1}}, \qquad n \in \bbZ,
\end{equation*}
and $q$-factorial
\begin{equation*}
[n]_q! = \prod_{m = 1}^n [m]_q, \quad n \in \bbN, \qquad [0]_q! = 1.
\end{equation*}
To simplify formulas we also denote
\begin{equation*}
\kappa_q = q - q^{-1}.
\end{equation*}

\section{\texorpdfstring{Quantum loop algebra $\uqlsllpo$}{Quantum loop 
algebra uqLsllpo}} \label{s:qla}

We start with reminding of basic definitions and fixing notation for the main
object of our interest, the quantum group of the loop algebra of the special 
linear Lie algebra of arbitrary rank. One usually denotes this object by 
$\uqlsllpo$ and calls it shortly the quantum loop algebra. Here, we also
recall two different realizations of this quantum loop algebra and relate
the corresponding generators explicitly.

\subsection{Drinfeld--Jimbo definition and Cartan--Weyl data} \label{s:cwg}

We fix a positive integer $l$ and introduce two sets $I = \{1, \ldots,l\}$ and 
$\hi = \{0, 1, \ldots,l\}$. Let $h_i$, $i \in I$, be the canonical Cartan generators 
of the Lie algebra $\sllpo$. Denote by $\gothh$ the canonical Cartan subalgebra of 
$\sllpo$, so that
\begin{equation*}
\gothh = \bigoplus_{i \in I} \bbC \, h_i.
\end{equation*}
The simple roots $\alpha_i$, $i \in I$, of $\sllpo$ are defined by the relations
\begin{equation*}
\langle \alpha_i, \, h_j \rangle = a_{j i}, \qquad i, \, j \in I,
\end{equation*}
where $a_{i j}$, $i, \, j \in I$, are the entries of the Cartan matrix of $\sllpo$. 
The full system $\triangle_+$ of positive roots of $\sllpo$ is
\begin{equation*}
\triangle_+ = \{ \alpha_{i j} \mid 1 \le i < j \le l + 1 \},
\end{equation*}
where
\begin{equation*}
\alpha_{i j} = \sum_{k = i}^{j - 1} \alpha_k.
\end{equation*}
It is clear that, in particular,
\begin{equation*}
\alpha_i = \alpha_{i, \, i + 1}.
\end{equation*}
Certainly, the system of negative roots is $\triangle_- = - \triangle_+$, and the full 
system of roots is $\triangle = \triangle_+ \sqcup \triangle_-$.

Following Kac, we use the notation $\lsllpo$ for the loop algebra of $\sllpo$, $\tlsllpo$ 
for its standard extension by a one-dimensional center $\bbC \, c$, and $\hlsllpo$ for the 
Lie algebra obtained from $\tlsllpo$ by adding a natural derivation $d$ \cite{Kac90}.

The Cartan subalgebra $\hgothh$ of $\hlsllpo$ is
\begin{equation*}
\hgothh = \gothh \oplus \bbC \, c \oplus \bbC \, d.
\end{equation*}
Introducing an additional Kac--Moody generator
\begin{equation*}
h_0 = c - \sum_{i \in I} h_i
\end{equation*}
we obtain
\begin{equation*}
\hgothh = \bigoplus_{i \in \hi} \bbC \, h_i + \bbC \, d.
\end{equation*}
It is worth to note that
\begin{equation*}
c = h_0 + \sum_{i \in I} h_i = \sum_{i \in \widehat I} h_i.
\end{equation*}
We identify the space $\gothh^*$ with the subspace of $\widehat \gothh^*$ defined as
\begin{equation*}
\gothh^* = \{ \widehat \gamma \in \widehat \gothh^* \mid \langle \widehat \gamma, \, c \rangle = 0, 
\quad \langle \widehat \gamma, \, d \rangle = 0 \}.
\end{equation*}
It is also convenient to denote
\begin{equation*}
\tgothh = \gothh \oplus \bbC \, c = \Bigl( \bigoplus_{i \in I} \bbC \, h_i \Bigr) \oplus \bbC \, c 
= \bigoplus_{i \in \widehat I} \bbC \, h_i
\end{equation*}
and identify the space $\gothh^*$ with the subspace of $\widetilde \gothh^*$ which consists of the elements 
$\widetilde \gamma \in \widetilde \gothh^*$ satisfying the condition
\begin{equation}
\langle \widetilde \gamma, \, c \rangle = 0. 
\label{lambdac}
\end{equation}
Here and everywhere below we mark such elements by a tilde. Explicitly the identification is performed 
as follows. The element $\widetilde \gamma \in \widetilde \gothh^*$ satisfying (\ref{lambdac}) is identified 
with the element $\gamma \in \gothh^*$ defined by the equations
\begin{equation*}
\langle \gamma, \, h_i \rangle = \langle \widetilde \gamma, \, h_i \rangle, \qquad i \in I.
\end{equation*}
In the opposite direction, given an element $\gamma \in \gothh^*$, we identify it with the element 
$\widetilde \gamma \in \widetilde \gothh^*$ determined by the relations
\begin{equation*}
\langle \widetilde \gamma, \, h_0 \rangle = - \sum_{i \in I} \langle \gamma, \, h_i \rangle, \qquad  
\langle \widetilde \gamma, \, h_i \rangle = \langle \gamma, \, h_i \rangle, \quad i \in I.
\end{equation*}
It is clear that $\widetilde \gamma$ satisfies (\ref{lambdac}).

The simple roots $\alpha_i \in \hgothh^*$, $i \in \hi$, of the Lie algebra $\hlsllpo$ are defined by the relations
\begin{equation*}
\langle \alpha_i, \, h_j \rangle = a_{j i}, \quad i, j \in \widehat I, \qquad \langle \alpha_0, \, d \rangle = 1, 
\qquad \langle \alpha_i, \, d \rangle = 0, \quad i \in I.
\end{equation*}
Here $a_{i j}$, $i, j \in \hi$, are the entries of the extended Cartan matrix of $\sllpo$. The full system $\widehat \triangle_+$ of positive roots of the Lie algebra $\hlsllpo$ is related to the system $\triangle_+$ of positive roots 
of $\sllpo$ as
\begin{multline*}
\widehat \triangle_+ = \{\gamma + n \delta \mid  \gamma \in \triangle_+, \ n \in \bbZ_+\} \\
\cup \{n \delta \mid n \in \bbN\} \cup \{(\delta - \gamma) + n \delta \mid  \gamma \in \triangle_+, \ n \in \bbZ_+\},
\end{multline*}
where
\begin{equation*}
\delta = \sum_{i \in \hi} \alpha_i
\end{equation*}
is the minimal positive imaginary root. 
It is worth to note here that
\begin{equation*}
\alpha_0 = \delta - \sum_{i \in I} \alpha_i = \delta - \theta,
\end{equation*}
where $\theta$ is the highest root of $\sllpo$. The system of negative roots $\widehat \triangle_-$ is $\widehat \triangle_- = - \widehat \triangle_+$, and we the full system of roots is
\begin{equation*}
\widehat \triangle = \widehat \triangle_+ \sqcup \widehat \triangle_- 
= \{ \gamma + n \delta \mid \gamma \in \triangle, \ n \in \bbZ \} \cup \{n \delta \mid n \in \bbZ \setminus \{0\} \}.
\end{equation*} 
Note that the set formed by the restriction of the simple roots $\alpha_i$ to $\tgothh$ is linearly dependent. 
In fact, we have
\begin{equation*}
\delta|_{\tgothh} = \sum_{i \in \hi} \alpha_i|_{\tgothh} = 0.
\end{equation*}
This is the main reason to pass from $\tlsllpo$ to $\hlsllpo$.

We fix a non-degenerate symmetric bilinear form on $\hgothh$ by the equations
\begin{equation*}
(h_i \mid h_j) = a^{\mathstrut}_{i j}, \qquad (h_i \mid d) = \delta_{i 0}, 
\qquad (d \mid d) = 0,
\end{equation*}
where $i, j \in \hi$. Then, for the corresponding symmetric bilinear form on 
$\hgothh^*$ one has
\begin{equation*}
(\alpha_i \mid \alpha_j) = a_{i j}.
\end{equation*}
It follows from this relation that
\begin{equation*}
(\delta \mid \alpha_{i j}) = 0, \qquad (\delta \mid \delta) = 0
\end{equation*}
for all $1 \le i < j \le l + 1$. 

To properly define the quantum loop algebra $\uqlsllpo$, we first introduce 
the quantum group $\uqhlsllpo$. It is a unital associative $\bbC$-algebra being a $q$-deformation of the universal enveloping algebra $\uhlsllpo$. It is generated by 
the elements $e_i$, $f_i$, 
$i \in \hi$, and $q^x$, $x \in \hgothh$, subject to the 
defining relations
\begin{gather}
q^0 = 1, \qquad q^{x_1} q^{x_2} = q^{x_1 + x_2}, \label{djra} \\
q^x e_i \, q^{-x} = q^{\langle \alpha_i, \, x \rangle} e_i, \qquad q^x f_i \, q^{-x} 
= q^{- \langle \alpha_i, \, x \rangle} f_i, \\
[e_i, \, f_j] = \delta_{i j} \, \frac{q^{h_i} - q^{- h_i}}{q^{\mathstrut} - q^{-1}}, \\
\sum_{k = 0}^{1 - a_{i j}} (-1)^k e_i^{(1 - a_{i j} - k)} e^{\mathstrut}_j \, e_i^{(k)} = 0, \qquad 
\sum_{k = 0}^{1 - a_{i j}} (-1)^k f_i^{(1 - a_{i j} - k)} f^{\mathstrut}_j \, f_i^{(k)} = 0,
\label{djrd}
\end{gather}
where $e_i^{(n)} = e_i^n / [n]_{q}!$, \, $f_i^{(n)} = f_i^n / [n]_{q}!$, and it is assumed 
that $i$ and $j$ are distinct in the Serre relations (\ref{djrd}).

The quantum group $\uqhlsllpo$ possesses no finite-dimensional representations with a 
nontrivial action of the element $q^{\nu c}$ \cite{ChaPre91, ChaPre94}. In contrast, 
the quantum loop algebra $\uqlsllpo$ has, along with the infinite-dimensional 
representations, also nontrivial finite-dimensional representations. Therefore, 
we proceed in two steps to the quantum loop algebra $\uqlsllpo$. As the first step we define 
the quantum group $\uqtlsllpo$ as a Hopf subalgebra of $\uqhlsllpo$ generated by the elements 
$e_i$, $f_i$, $i \in \hi$, and $q^x$, $x \in \tgothh$, with relations (\ref{djra})--(\ref{djrd}). 
Then, the quantum loop algebra $\uqlsllpo$ is defined as the quotient algebra of $\uqtlsllpo$ by 
the two-sided Hopf ideal generated by the elements of the form $q^{\nu c} - 1$ with $\nu \in \bbC^\times$. 
It is convenient to treat the quantum loop algebra $\uqlsllpo$ as a unital associative $\bbC$-algebra 
generated by the same generators as $\uqtlsllpo$ with relations (\ref{djra})--(\ref{djrd}) supplemented 
with the relations
\begin{equation}
q^{\nu c} = 1, \qquad \nu \in \bbC^\times. \label{qnuc}
\end{equation}
For the study of quantum integrable systems, it is important that the quantum loop 
algebra $\uqlsllpo$ is a Hopf algebra with appropriately defined co-multiplication, 
antipode and counit. However, the explicit form of the Hopf algebra structure is not 
used in the present paper, hence we omit it.

The abelian group
\begin{equation*}
\widehat Q = \bigoplus_{i \in \widehat I} \bbZ \, \alpha_i
\end{equation*}
is called the root lattice of $\hlsllpo$. The algebra $\uqlsllpo$ can be considered as $\widehat Q$-graded 
if we assume that
\begin{equation*}
e_i \in \uqlsllpo_{\alpha_i}, \qquad f_i \in \uqlsllpo_{- \alpha_i}, \qquad q^x \in \uqlsllpo_0
\end{equation*}
for any $i \in \widehat I$ and $x \in \tgothh$. An element $a$ of $\uqlsllpo$ is called a root vector corresponding to a root $\gamma$ of $\hlsllpo$ if $a \in \uqlsllpo_\gamma$. In particular, the generators $e_i$ and $f_i$ are root vectors corresponding to the roots $\alpha_i$ and $- \alpha_i$.

One can obtain linearly independent root vectors corresponding to all roots from $\widehat \triangle$. This can be done in many ways. We follow here the procedure proposed by Khoroshkin and Tolstoy \cite{TolKho92, KhoTol93}. Its main advantage is that it gives the root vectors closely related to the generators of Drinfeld's second realization of the quantum groups, see section \ref{s:2dr}. For an equivalent alternative, we also refer to the paper \cite{Bec94a}. The root vectors, together with the elements $q^x$, $x \in \widetilde \gothh$, are the Cartan--Weyl generators of $\uqlsllpo$. Appropriately ordered monomials constructed from these generators form a Poincar\'e--Birkhoff--Witt basis of $\uqlsllpo$. 

Let us describe how the procedure of Khoroshkin and Tolstoy looks in the case under consideration. First of all one assumes that a normal order $\prec$ on the system of 
positive roots $\widehat \triangle_+$ is given \cite{AshSmiTol79, Tol89}. For the case of a 
finite-dimensional simple Lie algebra it means that if a positive root $\gamma$ is a sum of 
two positive roots $\alpha$ and $\beta$, then either $\alpha \prec \gamma \prec \beta$ or $\beta \prec \gamma \prec \alpha$. In our case we assume additionally that 
\begin{equation}
\alpha + k \delta \prec m \delta \prec (\delta - \beta) + n \delta \label{akd}
\end{equation}
for any $\alpha, \beta \in \triangle_+$ and $k, m, n \in \bbZ_+$. We use in this paper the following normal order, 
see, for example \cite{MenTes15}. As is seen from (\ref{akd}) it suffices to define separately the ordering of the 
roots $\alpha + k \delta$ and $(\delta - \beta) + n \delta$, where $\alpha, \beta \in \triangle_+$. We assume that $\alpha_{i j} + r \delta \prec \alpha_{m n} + s \delta$ if $i < m$, or if $i = m$ and $r < s$, or if $i = m$, 
$r = s$ and $j < n$. In a similar way, $(\delta - \alpha_{i j}) + r \delta \prec (\delta - \alpha_{m n}) + s \delta$ 
if $i > m$, or if $i = m$ and $r > s$, or if $i = m$, $r = s$ and $j < n$.

The root vectors are defined by the following inductive procedure. We take as the 
root vectors, corresponding to the simple roots the corresponding generators of 
$\uqlsllpo$,
\begin{equation*}
e_{\delta - \theta} = e_0, \qquad e_{\alpha_i} = e_i, \qquad f_{\delta - \theta} = f_0, 
\qquad f_{\alpha_i} = f_i, \qquad i \in I.
\end{equation*}
Here and below we denote a root vector corresponding to a positive root $\gamma$ by 
$e_\gamma$, and a root vector corresponding to a negative root $- \gamma$ by $f_\gamma$.

Let now a root $\gamma \in \widehat \triangle_+$ be such that $\gamma = \alpha + \beta$ for some 
$\alpha, \beta \in \widehat \triangle_+$. For definiteness assume that $\alpha \prec \gamma \prec \beta$. 
Further, let there be no other roots $\alpha' \succ \alpha$ and $\beta' \prec \beta$ satisfying 
$\gamma = \alpha' + \beta'$. Then, if the root vectors $e_\alpha$, $e_\beta$ and $f_\alpha$, $f_\beta$ 
are already given, we define 
\cite{TolKho92, KhoTol93}
\begin{equation*}
e_\gamma = [e_\alpha \, , \, e_\beta]_q, \qquad 
f_\gamma = [f_\beta \, , \, f_\alpha]_q.
\label{e1}
\end{equation*}
Here we have introduced the $q$-commutator $[ \, \ , \ ]_q$ defined as
\begin{equation*}
[e_\alpha \, , \, e_\beta]_q 
= e_\alpha \, e_\beta - q^{-(\alpha \, | \, \beta)} \, e_\beta \, e_\alpha, \qquad
[f_\alpha \, , \, f_\beta]_q 
= f_\alpha \, f_\beta - q^{(\alpha \, | \, \beta)} \, f_\beta \, f_\alpha,
\end{equation*}
where $( \, | \, )$ denotes the symmetric bilinear form on $\widehat \gothh^*$.

First we define root vectors corresponding to the roots $\alpha_{i j}$ and $- \alpha_{i j}$. Recall that 
we already have root vectors $e_{\alpha_{i, \, i + 1}}$ and $f_{\alpha_{i, \, i + 1}}$ corresponding to 
the roots $\alpha_{i, i + 1} = \alpha_i$ and $- \alpha_{i, i + 1} = - \alpha_i$. We define root vectors 
corresponding to the positive composite roots by the inductive rule
\begin{equation*}
e_{\alpha_{i j}} = [e_{\alpha_{i, \, i + 1}}, \, e_{\alpha_{i + 1, \, j}}]_q 
= e_{\alpha_{i, \, i + 1}} e_{\alpha_{i + 1, \, j}} - q \, e_{\alpha_{i + 1, \, j}} e_{\alpha_{i, \, i + 1}},
\end{equation*}
and similarly for the negative roots
\begin{equation*}
f_{\alpha_{i j}} = [f_{\alpha_{i + 1, \, j}}, \, f_{\alpha_{i, \, i + 1}}]_q 
= f_{\alpha_{i + 1, \, j}} f_{\alpha_{i, \, i + 1}} - q^{-1} f_{\alpha_{i, \, i + 1}} f_{\alpha_{i + 1, \, j}}.
\label{faij}
\end{equation*}
It is clear that these rules uniquely give
\begin{equation*}
e_{\alpha_{i j}} = [e_{\alpha_i}, \, \ldots \, [e_{\alpha_{j - 2}}, \, e_{\alpha_{j - 1}} \, ]_q \, \ldots \, ]_q, 
\qquad
f_{\alpha_{i j}} = [ \, \ldots \, [f_{\alpha_{j - 1}}, \, f_{\alpha_{j - 2}}]_q \, \ldots, \, f_{\alpha_i}]_q.
\end{equation*}
Roughly speaking, we start with the simple root $\alpha_{j - 1} = \alpha_{j - 1, j}$ 
and then sequentially add necessary simple roots from the left to reach the final root $\alpha_{i j}$. In fact, we can start with any simple root $\alpha_k$ with $i < k < j$ 
and then add the appropriate simple roots from the left or from the right in a random 
order. It appears that the resulting root vector will be the same.

Now we consider the roots of the form $\delta - \alpha_{i j}$ and $- (\delta - \alpha_{i j})$. The root vectors $e_{\delta - \theta}$ and $f_{\delta - \theta}$ corresponding to the roots $\delta - \theta = \delta - \alpha_{1, \, l + 1}$ and $-(\delta - \theta) = - (\delta - \alpha_{1, \, l + 1})$ are already given. We define inductively
\begin{align}
& e_{\delta - \alpha_{i j}} = [e_{\alpha_{i - 1, \, i}}, \,  e_{\delta - \alpha_{i - 1, \, j}}]_q 
= e_{\alpha_{i - 1, \, i}} \,  e_{\delta - \alpha_{i - 1, \, j}} - q \, e_{\delta - \alpha_{i - 1, \, j}} \, 
e_{\alpha_{i - 1, \, i}}, \label{edmga} \\
& f_{\delta - \alpha_{i j}} = [f_{\delta - \alpha_{i - 1, \, j}}, \, f_{\alpha_{i - 1, \, i}}]_q 
= f_{\delta - \alpha_{i - 1, \, j}} \, f_{\alpha_{i - 1, \, i}} - q^{-1} f_{\alpha_{i - 1, \, i}} \, 
f_{\delta - \alpha_{i - 1, \, j}} \label{edmgb}
\end{align}
if $i > 1$, and
\begin{align}
& e_{\delta - \alpha_{1 j}} = [e_{\alpha_{j, \, j + 1}}, \,  
e_{\delta - \alpha_{1, \, j + 1}}]_q  = e_{\alpha_{j, \, j + 1}} \,  
e_{\delta - \alpha_{1, \, j + 1}} 
- q \, e_{\delta - \alpha_{1, \, j + 1}} \,  e_{\alpha_{j, \, j + 1}}, \label{edmgc} \\
& f_{\delta - \alpha_{1 j}} = [f_{\delta - \alpha_{1, \, j + 1}}, \, 
f_{\alpha_{j, \, j + 1}}]_q  =  f_{\delta - \alpha_{1, \, j + 1}} \, f_{\alpha_{j, \, j + 1}} 
- q^{-1} f_{\alpha_{j, \, j + 1}} \, f_{\delta - \alpha_{1, \, j + 1}} \label{edmgd}
\end{align}
for $j < l + 1$. The inductive rules (\ref{edmga}), (\ref{edmgb}) and (\ref{edmgc}), (\ref{edmgd}) 
uniquely lead to the expressions 
\begin{align}
e_{\delta - \alpha_{i j}} & = [e_{\alpha_{i-1}}\, , \, \ldots [e_{\alpha_1} \, , \, 
[e_{\alpha_{j}} \, , \ldots [e_{\alpha_l}\, , \, e_{\delta - \theta}]_q \ldots ]_q \, ]_q
\ldots ]_q, \label{edaij} \\[.3em]
f_{\delta - \alpha_{i j}} & = [\ldots [ \, [ \ldots [ f_{\delta - \theta} \, , \, f_{\alpha_l} ]_q \, , \ldots f_{\alpha_{j}} ]_q \, , \, f_{\alpha_1} \, ]_q \, , \ldots f_{\alpha_{i - 1}} ]_q. \label{fdaij}
\end{align}
Saying in words, we start with the highest root $\theta$ and then sequentially subtract redundant simple roots 
first from the right and then from the left. We can randomly interchange subtractions from the left and from the 
right, the result will be the same.

Then, assuming that $j = i + 1$, so that $\alpha_{i j}$ at the left hand side of 
relations (\ref{edaij}), (\ref{fdaij}) reduces to any of the simple roots $\alpha_i$, 
$i \in I$, we obtain
\begin{align}
e_{\delta - \alpha_1} & = [ e_{\alpha_2} \, , \, [ e_{\alpha_3} \, , \ldots [e_{\alpha_l} 
\, , \, e_{\delta - \theta}]_q \ldots ]_q \, ]_q, \label{edma1} \\[.3em]
e_{\delta - \alpha_i} & = [e_{\alpha_{i-1}}\, , \, \ldots [e_{\alpha_1} \, , \, 
[e_{\alpha_{i+1}} \, , \ldots [e_{\alpha_l}\, , \, e_{\delta - \theta}]_q \ldots ]_q \, ]_q
\ldots ]_q, \quad i = 2, \ldots, l-1, \label{edmai} \\[.3em]
e_{\delta - \alpha_l} & = [e_{\alpha_{l-1}} \, , \, \ldots [e_{\alpha_2}\, , \, 
[e_{\alpha_1} \, , \, e_{\delta - \theta}]_q]_q \ldots ]_q \label{edmal}
\end{align}
and, similarly,
\begin{align}
f_{\delta - \alpha_1} & = [ \, [ \ldots [ f_{\delta - \theta} \, , \, f_{\alpha_l} ]_q \, , 
\ldots f_{\alpha_3} ]_q \, , f_{\alpha_2} ]_q, \label{fdma1} \\[.3em]
f_{\delta - \alpha_i} & = [\ldots [ \, [ \ldots [ f_{\delta - \theta} \, , \, f_{\alpha_l} ]_q
\, , \ldots f_{\alpha_{i+1}} ]_q \, , \, f_{\alpha_1} \, ]_q \, , \ldots f_{\alpha_{i-1}} ]_q, \quad i = 2, 
\ldots, l - 1, \label{fdmai} \\[.3em]
f_{\delta - \alpha_l} & = [ \ldots [ \, [ f_{\delta - \theta} \, , f_{\alpha_1} ]_q \, , \,
f_{\alpha_2} ]_q \, , \ldots f_{\alpha_{l - 1}} ]_q. \label{fdmal}
\end{align}

Having all root vectors corresponding to the roots $\alpha_{i j}$ and $\delta - \alpha_{i j}$ with 
$\alpha_{i j} \in \triangle_+$, we can continue by adding imaginary roots $n \delta$. 
The root vectors corresponding to the imaginary roots are additionally indexed by the 
positive roots $\gamma \in \triangle_+$ of $\sllpo$ and are defined by the relations
\begin{equation}
e'_{\delta, \, \gamma} = [e_\gamma \, , \, e_{\delta - \gamma}]_q, 
\qquad f'_{\delta, \, \gamma} = [f_{\delta - \gamma} \, , \, f_{\gamma}]_q. 
\label{epd}
\end{equation}
Then, the remaining higher root vectors are given by the iteration 
\cite{TolKho92, KhoTol93}
\begin{gather}
e_{\gamma + n \delta} = ([2]_q)^{-1} 
[e_{\gamma + (n - 1)\delta} \, , \, e'_{\delta, \, \gamma}]_q, \qquad 
f_{\gamma + n \delta} = ([2]_q)^{-1} 
[f'_{\delta, \, \gamma} \, , \, f_{\gamma + (n - 1)\delta}]_q, \label{cwbx} \\
e_{(\delta - \gamma) + n \delta} = ([2]_q)^{-1} 
[e'_{\delta, \, \gamma} \, , \, e_{(\delta - \gamma) + (n - 1)\delta}]_q, \label{cwby1} \\
f_{(\delta - \gamma) + n \delta} = ([2]_q)^{-1} 
[f_{(\delta - \gamma) + (n - 1)\delta} \, , \, f'_{\delta, \, \gamma}]_q, \label{cwby2} \\
e'_{n \delta, \, \gamma} = [e_{\gamma + (n - 1)\delta} \, , \, e_{\delta - \gamma}]_q, 
\qquad f'_{n \delta, \, \gamma} = [f_{\delta - \gamma} \, , \, f_{\gamma + (n - 1)\delta}]_q, \label{cwbz}
\end{gather}
where we take into account that $(\alpha_{i j} \, | \, \alpha_{i j}) = 2$ for all 
permissible values of $i$ and $j$. We note that, among all imaginary root vectors 
$e'_{n\delta, \, \gamma}$ and $f'_{n\delta, \, \gamma}$ only the root vectors 
$e'_{n \delta, \, \alpha_i}$ and $f'_{n \delta, \, \alpha_i}$, $i \in I$, 
are independent and required for the construction of the Poincar\'e--Birkhoff--Witt 
basis. However, the vectors $e'_{\delta, \gamma}$ and $f'_{\delta, \gamma}$ with arbitrary 
$\gamma \in \triangle_+$ are needed for the iterations (\ref{cwby1}) and (\ref{cwby2}).

The prime in the notation for the root vectors corresponding to the imaginary 
roots $n \delta$ and $- n \delta$, $n \in \bbN$ is justified by the fact that 
one also uses another set of root vectors corresponding to these roots. They 
are introduced by the functional equations
\begin{gather}
- \kappa_q \, e_{\delta, \gamma}(u) = \log(1 - \kappa_q \, e'_{\delta, \, \gamma}(u)), \label{edg} \\
\kappa_q \, f_{\delta, \gamma}(u^{-1}) = \log(1 + \kappa_q \, f'_{\delta, \, \gamma}(u^{-1})), \label{fdg}
\end{gather}
where the generating functions
\begin{align*}
& e'_{\delta, \, \gamma}(u) = \sum_{n = 1}^\infty e'_{n \delta, \, \gamma} \, u^n, && e_{\delta, \, \gamma}(u) 
= \sum_{n = 1}^\infty e_{n \delta, \, \gamma} \, u^n, \\
& f'_{\delta, \, \gamma}(u^{-1}) = \sum_{n = 1}^\infty f'_{n \delta, \, \gamma} \, u^{- n}, && f_{\delta, \, \gamma}(u^{-1}) = \sum_{n = 1}^\infty f_{n \delta, \, \gamma} \, u^{- n}
\end{align*}
are defined as formal power series. In particular, it is convenient to use the unprimed imaginary root vectors for the universal $R$-matrix of quantum affine algebras 
\cite{TolKho92, KhoTol93}.

\subsection{Drinfeld's second realization} \label{s:2dr}

The quantum loop algebra $\uqlsllpo$ can be realized in a different way \cite{Dri87, Dri88} 
as a $\bbC$-algebra with generators $\xi^\pm_{i, \, n}$ with $i \in I$ and $n \in \bbZ$, 
$q^x$ with $x \in \gothh$, and $\chi_{i, \, n}$ with $i \in I$ and $n \in \bbZ \setminus \{0\}$. 
They satisfy the defining relations
\begin{gather*}
q^0 = 1, \qquad q^{x_1} q^{x_2} = q^{x_1 + x_2}, \\
[\chi^{\mathstrut}_{i, \, n}, \, \chi^{\mathstrut}_{j, \, m}] = 0, \qquad 
q^x \chi_{j, \, n} = \chi_{j, \, n} \, q^x,  \\
q^x \xi^\pm_{i, \, n} q^{- x} = q^{\pm \langle \alpha_i, \, x \rangle} \xi^\pm_{i, \, n}, 
\qquad [\chi^{\mathstrut}_{i, \, n}, \, \xi^\pm_{j, m}] = \pm \frac{1}{n} 
[n \, a_{i j}]^{\mathstrut}_{q} \, \xi^\pm_{j, \, n + m}, \\
\xi^\pm_{i, \, n + 1} \xi^\pm_{j, \, m} - q^{\pm a_{i j}} \, \xi^\pm_{j, \, m} \, 
\xi^\pm_{i, \, n + 1} = q^{\pm a_{i j}} \, \xi^\pm_{i, \, n} \, 
\xi^\pm_{j, \, m + 1} - \xi^\pm_{j, \, m + 1} \xi^\pm_{i, \, n}, \\
[\xi^+_{i, \, n}, \, \xi^-_{j, \, m}] = \delta_{i j} \, \frac{\phi^+_{i, \, n + m} 
- \phi^-_{i, \, n + m}}{q^{\mathstrut} - q^{-1}}
\end{gather*}
and the Serre relations whose explicit form is not important for our consideration. 
In this realization, the entries of the generalized Cartan matrix $a_{i j}$ of type 
$A_l$ take place. The quantities $\phi^\pm_{i, \, n}$, $i \in I$, $n \in \bbZ$, are 
given by the formal power series
\begin{equation}
\sum_{n = 0}^\infty \phi^\pm_{i, \, \pm n} u^{\pm n} = q^{\pm h_i} 
\exp \left( \pm \kappa_q \sum_{n = 1}^\infty \chi_{i, \, \pm n} u^{\pm n} \right) \label{phipm}
\end{equation}
supplemented with the conditions
\begin{equation*}
\phi^+_{i, \, n} = 0, \quad n < 0, \qquad \phi^-_{i, \, n} = 0, \quad n > 0.
\end{equation*}

We see that each of the generators $\xi^+_{i, \, n}$, $\xi^-_{i, \, n}$ is labeled 
by two integers, $i \in I$ and $n \in \bbZ$, exactly as each of the root vectors 
$e_{\alpha_i + n\delta}$, $f_{\alpha_i + n\delta}$ and $e_{\delta - \alpha_i +n\delta}$,
$f_{\delta - \alpha_i + n\delta}$. Further, each of the generators of the commutative 
subalgebra $\chi_{i, \, n}$, $\chi_{i, \, -n}$, or equivalently $\phi^+_{i, \, n}$, 
$\phi^-_{i, \, -n}$ just in view of (\ref{phipm}), is labeled by two integers $i \in I$ 
and $n \in \bbN$, exactly as the imaginary root vectors $e_{n\delta, \, \alpha_i}$, 
$f_{n\delta, \, \alpha_i}$, or equivalently $e'_{n\delta, \, \alpha_i}$, 
$f'_{n\delta, \, \alpha_i}$ just in view of (\ref{edg}), (\ref{fdg}). 
Indeed, there is an isomorphism of the two realizations of the quantum 
loop algebras which can be expressed by a one-to-one correspondence 
between the above generators and root vectors.

To be precise, the generators of Drinfeld's second realization are related to the 
Cartan--Weyl generators in the following way \cite{KhoTol93, KhoTol94}. The generators 
$q^x$ of the quantum loop algebra in Drinfeld--Jimbo's and Drinfeld's second 
realizations are the same, except that in the first case $x \in \widetilde \gothh$, 
while in the second case $x \in \gothh \subset \widetilde \gothh$. For the generators 
$\xi^\pm_{i, \, n}$ and $\chi_{i, \, n}$ of Drinfeld's second realization we have
\begin{gather}
\xi^+_{i, \, n} = \begin{cases}
(-1)^n o_i^n e_{\alpha_i + n \delta} & n \ge 0 \\
(-1)^{n + 1} o_i^n q^{-h_i} f_{(\delta - \alpha_i) - (n + 1)\delta} & n < 0
\end{cases}, \label{ksipin} \\
\xi^-_{i, \, n} = \begin{cases}
(-1)^n o_i^{n + 1} e_{(\delta - \alpha_i) + (n - 1) \delta} \, q^{h_i} & n > 0 \\
(-1)^n o_i^n f_{\alpha_i - n \delta} & n \le 0
\end{cases}, \label{ksimin} \\
\chi_{i, \, n} = \begin{cases}
(-1)^{n + 1} o_i^n e_{n \delta, \, \alpha_i} & n > 0 \\
(-1)^{n + 1} o_i^n f_{- n\delta, \, \alpha_i} & n < 0
\end{cases}, \label{chiin}
\end{gather}
where for each $i \in I$ the number $o_i$ is either $+1$ or $-1$, such that $o_i = - o_j$ 
whenever $a_{i j} < 0$. It follows from (\ref{edg}), (\ref{fdg}), (\ref{phipm}) and (\ref{chiin}) 
that
\begin{gather*}
\phi^+_{i, \, n} = \begin{cases}
(-1)^{n + 1} o_i^n \kappa_q \, q^{h_i} e'_{n \delta, \, \alpha_i} & n > 0 \\
q^{h_i} & n = 0
\end{cases}, \\
\phi^-_{i, \, n} = \begin{cases}
q^{-h_i} & n = 0 \\
(-1)^n o_i^n \kappa_q \, q^{-h_i} f'_{- n \delta, \, \alpha_i} & n < 0
\end{cases}.
\end{gather*}
Defining the generating functions $\phi^+_i(u)$ and $\phi^-_i(u)$ as
\begin{equation*}
\phi^+_i(u) = \sum_{n = 0}^\infty \phi^+_{i, \, n} u^n, \qquad \phi^-_i(u^{-1}) 
= \sum_{n = 0}^\infty \phi^-_{i, \, -n} u^{- n},
\end{equation*}
we also obtain
\begin{align}
& \phi^+_i(u) = q^{h_i} \big(1 - \kappa_q \, e'_{\delta, \, \alpha_i}(- o_i u)\big), \label{phipiu} \\ 
& \phi^-_i(u^{-1}) = q^{-h_i} \big(1 + \kappa_q \, f'_{\delta, \, \alpha_i}(- o_i u^{-1})\big). \label{phimiu}
\end{align}
In the case under consideration, the numbers $o_i$ entering equations 
(\ref{phipiu}) and (\ref{phimiu}) can be explicitly defined as
\begin{equation}
o_i = - (-1)^i, \qquad i = 1, \ldots, l.
\label{oi}
\end{equation}
This specification perfectly matches the general definition.

\section{\texorpdfstring{Highest $\ell$-weight representations of Borel subalgebras}{Highest l-weight 
representations of Borel subalgebras}} 
\label{s:rbs}

We first recall some general information about the highest $\ell$-weight representations of the Borel subalgebras of the quantum loop algebra $\uqlsllpo$. Here we follow mainly the papers \cite{MukYou14, HerJim12}, where basic properties of modules over the quantum loop algebras and their Borel subalgebras in the category $\calO$ were systematically studied and presented in a most suitable for our purposes form. We also refer to \cite{BooGoeKluNirRaz16} for the consistency of the notations. Then we proceed to examples of the highest $\ell$-weight representations we are particularly interested in.

\subsection{Preliminaries} \label{s:gibs}

The quantum loop algebra $\uqlsllpo$ has two Borel subalgebras, the positive $\uqbp$ 
and the negative $\uqbm$ ones. As we have already mentioned, the universal$R$-matrix 
belongs to a completion of their tensor product. Therefore, representations of these Borel
subalgebras are exactly what one needs for applications in the theory of integrable systems. 
In terms of the Drinfeld--Jimbo generators, the Borel subalgebras are defined as follows. 
The Borel subalgebra $\uqbp$ is the subalgebra generated by $e_i$, $i \in \hi$, 
and $q^x$, $x \in \tgothh$, and the Borel subalgebra $\uqbm$ is the subalgebra generated 
by $f_i$, $i \in \hi$, and $q^x$, $x \in \tgothh$. It is clear that the Borel
subalgebras are Hopf subalgebras of $\uqlsllpo$. The description of $\uqbp$ and $\uqbm$ 
in terms of the Drinfeld generators is more intricate. We note that, as follows from  
(\ref{ksipin})--(\ref{chiin}), the Borel subalgebra $\uqbp$ contains the Drinfeld 
generators $\xi^+_{i, n}$, $\xi^-_{i, m}$, $\chi_{i, m}$ with $i \in I$, $n \geq 0$ 
and $m > 0$, while the Borel subalgebra $\uqbm$ contains the Drinfeld generators 
$\xi^-_{i, n}$, $\xi^+_{i, m}$, $\chi_{i, m}$ with $i \in I$, $n \leq 0$ and $m < 0$. 
The positive and negative Borel subalgebras are related by the quantum Chevalley 
involution. Therefore, we restrict ourselves to the consideration of the subalgebra 
$\uqbp$.

As is noted above we consider the $\uqbp$-modules in the category $\calO$. By definition, 
any $\uqbp$-module $W$ in the category $\calO$ allows for the weight decomposition
\begin{equation*}
W = \bigoplus_{\lambda \in \gothh^*}  W_\lambda, \qquad
W_\lambda = \{w \in W \mid q^x w = q^{\langle \widetilde \lambda, \, x \rangle} w 
\mbox{ for any } x \in \tgothh \},
\end{equation*}
where $W_\lambda$ is called the weight space of weight $\lambda$, and a nonzero
element of $W_\lambda$ is called a weight vector of weight $\lambda$. Further, 
a $\uqbp$-module $W$ in the category $\calO$ is called the highest weight module 
with highest weight $\lambda$ if there exists a weight vector $v \in W$ of weight 
$\lambda$ such that
\begin{equation*}
e_i \, v = 0
\end{equation*}
for all $i \in I$ and 
\begin{equation*}
W = \uqbp v.
\end{equation*}
The vector $v$ with such properties is called the highest weight vector of $W$.
It is uniquely defined up to a scalar factor. 

The notion of highest weight $\uqbp$-modules is refined by introducing $\ell$-weights. 
Relative to the Borel subalgebra $\uqbp$, an $\ell$-weight $\bm \Psi$ is a pair\footnote{The 
symbol $+$ means that we work with the Borel subalgebra $\uqbp$. For the case of $\uqbm$ one 
defines an $\ell$ weight as a pair $(\lambda, \, \bm \Psi^-)$ and for the case of the whole 
quantum loop algebra $\uqlsllpo$ as a triple $(\lambda, \, \bm \Psi^+, \, \bm \Psi^-)$, see
also \cite{Her05}.}
\begin{equation*}
\bm \Psi = (\lambda, \, \bm \Psi^+),
\end{equation*}
where $\lambda \in \gothh^*$ and $\bm \Psi^+$ is an $l$-tuple
\begin{equation*}
\bm \Psi^+ = (\Psi^+_i(u))_{i \in I}
\end{equation*}
of formal series
\begin{equation*}
\Psi_i^+(u) = \sum_{n \in \bbZ_+} \Psi^+_{i, \, n} \, u^n \in \bbC[[u]],
\end{equation*}
such that
\begin{equation*}
\Psi^+_{i, \, 0} = q^{\langle \lambda, \, h_i \rangle}.
\end{equation*}

For any $\uqbp$-module in the category $\calO$ one has the $\ell$-weight decomposition
\begin{equation*}
W = \bigoplus_{\bm \Psi} W_{\bm \Psi},
\end{equation*}
where $W_{\bm \Psi}$ is a subspace of $W$ such that for any $w$ in $W_{\bm \Psi}$ 
there is $p \in \bbN$ such that
\begin{equation}
(\phi^+_{i, \, n} - \Psi^+_{i, \, n})^p w = 0
\label{intp}
\end{equation}
for all $i \in I$ and $n \in \bbZ_+$, and
\begin{equation*}
q^x w = q^{\langle \widetilde \lambda, \, x \rangle} w
\end{equation*}
for all $x \in \tgothh$. The subspace $W_{\bm \Psi}$ is called the 
$\ell$-weight space of $\ell$-weight $\bm \Psi$, and one says that $\bm \Psi$ is an $\ell$-weight 
of $W$ if the subspace $W_{\bm \Psi}$ is not trivial. A nonzero element $v \in W_{\bm \Psi}$ such 
that
\begin{equation*}
\phi^+_{i, \, n} v = \Psi^+_{i, \, n} v
\end{equation*}
for all $i \in I$ and $n \in \bbZ_+$ is said to be an $\ell$-weight vector of $\ell$-weight $\bm \Psi$. 
Every nontrivial $\ell$-weight space contains an $\ell$-weight vector.

A $\uqbp$-module $W$ in the category $\calO$ is called a highest $\ell$-weight module with  highest $\ell$-weight 
$\bm \Psi$ if there exists an $\ell$-weight vector $v \in W$ of $\ell$-weight $\bm \Psi$ such that
\begin{equation*}
\xi^+_{i, \, n} v = 0
\end{equation*}
for all $i \in I$ and $n \in \bbZ_+$ and
\begin{equation*}
W = \uqbp v.
\end{equation*}
The vector with the above properties is unique up to a scalar factor. One calls it the 
highest $\ell$-weight vector of $W$.

An $\ell$-weight $\bm \Psi$ of a $\uqbp$-module is called rational if for some non-negative integers 
$p_i$, $q_i$, $i \in I$, and complex numbers $a_{i r}$, $b_{i s}$, 
$i \in I$, $0 \le r \le p_i$, $0 \le s \le q_i$, one has
\begin{equation*}
\Psi^+_i(u) = \frac{a_{i p_i} u^{p_i} + a_{i, \, p_i - 1} u^{p_i - 1} 
+ \cdots + a_{i 0}}{b_{i q_i} u^{q_i} + b_{i, \, q_i - 1} u^{q_i - 1} 
+ \cdots + b_{i 0}}.
\end{equation*}
Here, the numbers $a_{i 0}$, $b_{i 0}$ must be nonzero. We see that the definition
of rational $\ell$-weights in the case of Borel subalgebras is essentially less 
restrictive than that of rational $\ell$-weights for the full quantum loop algebras,
where one should additionally impose certain regularity conditions \cite{MukYou14, BooGoeKluNirRaz16}.

One can show that for any rational $\ell$-weight $\bm \Psi$ there is a simple highest $\ell$-weight 
$\uqbp$-module $L(\bm \Psi)$ with highest $\ell$-weight $\bm \Psi$ which 
is unique up to an isomorphism, and any simple $\uqbp$-module in the category $\calO$ 
is a highest $\ell$-weight module with a rational highest $\ell$-weight \cite{HerJim12}. 
Here, all $\ell$-weights of a $\uqbp$-module in the category $\calO$ are rational.

One defines the product of $\ell$-weights $\bm \Psi_1 = (\lambda_1, \, \bm \Psi_1^+)$ and 
$\bm \Psi_2 = (\lambda_2, \, \bm \Psi_2^+)$ as the pair
\begin{equation*}
\bm \Psi_1 \bm \Psi_2 = (\lambda_1 + \lambda_2, \, \bm \Psi^+_1 \bm \Psi^+_2),
\end{equation*}
where
\begin{equation*}
\bm \Psi_1^+ \bm \Psi_2^+ = (\Psi_{1 i}^+(u) \Psi_{2 i}^+(u))_{i \in I}.
\end{equation*}
For any rational $\ell$-weights $\bm \Psi_1$ and $\bm \Psi_2$ 
the submodule of $L(\bm \Psi_1) \otimes L(\bm \Psi_2)$ generated 
by the tensor product of the highest $\ell$-weight vectors is a highest $\ell$-weight module with highest 
$\ell$-weight $\bm \Psi_1 \bm \Psi_2$. In particular, 
$L(\bm \Psi_1 \bm \Psi_2)$ is isomorphic to a subquotient 
of $L(\bm \Psi_1) \otimes L(\bm \Psi_2)$  generated by the 
tensor product of the highest $\ell$-weight vectors. We denote this subquotient as 
$L(\bm \Psi_1) \ovotimes L(\bm \Psi_2)$. 
Note that the operation $\ovotimes$ is associative.

We have already noted in section \ref{s:cwg} a special role of the higher
root vectors $e'_{n \delta, \, \alpha_i}$ and $f'_{n \delta, \, \alpha_i}$. 
Besides, we see from equation (\ref{phipiu}) and the definition of the highest 
$\ell$-weight representations of $\uqbp$ that only the root vectors 
$e'_{n \delta, \, \alpha_i}$ are used to determine the corresponding 
highest $\ell$-weight vectors and highest $\ell$-weights.

\subsection{Prefundamental representations} \label{s:prefreps}

Prefundamental representations \cite{HerJim12} are simple highest $\ell$-weight 
$\uqbp$-modules $L^\pm_{i, \, a}$ with the highest $\ell$-weights 
$(\lambda^\ms_{i, a}, \, \bm (\Psi^\pm_{i, \, a})^+)$ determined by 
the relations
\begin{equation*}
\lambda^\ms_{i, a} = 0, \qquad \bm (\Psi^\pm_{i, \, a})^+(u) 
= ( \underbracket[.6pt]{1, \, \ldots, \, 1}_{i - 1}, \, (1 - a u)^{\pm 1}, 
\, \underbracket[.6pt]{1, \, \ldots, \, 1}_{l - i} ), \quad i \in I, \quad a \in \bbC^\times.
\end{equation*}
For any $\xi \in \gothh^*$ the one-dimensional representation with the highest $\ell$-weight 
$\bm \Psi_\xi = (\lambda^\ms_\xi, \, (\bm \Psi_\xi)^+)$ defined as 
\begin{equation}
\lambda_\xi = \xi, \qquad (\bm \Psi_\xi)^+ 
= (q^{\langle \xi, \, h_1 \rangle}, \, \ldots \, q^{\langle \xi, \, h_l \rangle}) \label{psixi}
\end{equation}
is also included into the class of the prefundamental representations. The corresponding $\uqbp$-module is denoted by $L_\xi$.

For any $\uqbp$-module $V$ in category $\calO$ and an element $\xi \in \gothh^*$, we define a shifted $\uqbp$-module $V[\xi]$ shifting the action of the generators $q^x$. Namely, if $\varphi$ is the representation of $\uqbp$ corresponding to the module $V$ and $\varphi[\xi]$ is the representation corresponding to the module $V[\xi]$, then
\begin{equation*}
\varphi[\xi](e_i) = \varphi(e_i), \quad i \in I, \qquad \varphi[\xi](q^x) = q^{\langle \widetilde \xi, 
\, x \rangle} \varphi(q^x), \quad x \in \tgothh.
\end{equation*}
It is clear that the module $V[\xi]$ is in category $\calO$ and is isomorphic to $V \otimes L_\xi$.

One can show that any $\uqbp$-module in the category $\calO$ is a subquotient of a tensor product of prefundamental representations.

\subsection{\texorpdfstring{$q$-oscillator modules in general}{q-oscillator 
modules in general}}

To obtain a representation of a Borel subalgebra one can simply take the restriction of a representation of the full quantum loop algebra to this subalgebra. However,  for the theory 
of integrable systems more representations are needed. Here one constructs necessary representations first defining a homomorphism of a Borel subalgebra to the $q$-oscillator algebra or to the tensor product of several copies of this algebra. Then one uses the appropriate representations of the $q$-oscillator algebras and comes to the desirable representation of the Borel subalgebra. In this section we give the definition of the $q$-oscillator algebra and 
describe its important representations.

The $q$-oscillator algebra $\Osc_q$ is a unital associative $\bbC$-algebra with 
generators $b^\dagger$, $b$, $q^{\nu N}$, $\nu \in \bbC$, and relations
\begin{gather*}
q^0 = 1, \qquad q^{\nu_1 N} q^{\nu_2 N} = q^{(\nu_1 + \nu_2)N}, \\
q^{\nu N} b^\dagger q^{-\nu N} = q^\nu b^\dagger, \qquad 
q^{\nu N} b q^{-\nu N} = q^{-\nu} b, \\
b^\dagger b = \frac{q^N - q^{- N}}{q - q^{-1}}, \qquad 
b b^\dagger = \frac{q q^N - q^{-1} q^{- N}}{q - q^{-1}}.
\end{gather*}
Two representations of $\Osc_q$ are interesting for us. First, let $W^{\scriptscriptstyle +}$ be the free vector space generated by the set $\{ v_0, v_1, \ldots \}$. One can show that the relations
\begin{gather*}
q^{\nu N} v_m = q^{\nu m} v_m,  \\[.3em]
b^\dagger v_m = v_{m + 1}, \qquad b \, v_m = [m]_q v_{m - 1},
\end{gather*}
where we assume that $v_{-1} = 0$, endow $W^{\scriptscriptstyle +}$ with the structure of an $\Osc_q$-module. 
We denote the corresponding representation of the algebra $\Osc_q$ by $\chi^{\scriptscriptstyle +}$. Further, 
let $W^{\scriptscriptstyle -}$ be the free vector space generated again by the set $\{ v_0, v_1, \ldots \}$. 
The relations
\begin{gather*}
q^{\nu N} v_m = q^{- \nu (m + 1)} v_m,  \\[.3em]
b \, v_m = v_{m + 1}, \qquad b^\dagger v_m = - [m]_q v_{m - 1},
\end{gather*}
where we again assume that $v_{-1} = 0$, endow the vector space $W^{\scriptscriptstyle -}$ with the structure 
of an $\Osc_q$-module. We denote the corresponding representation of $\Osc_q$ by $\chi^{\scriptscriptstyle -}$.

We consider the tensor product of $l$ copies of the $q$-oscillator algebra, 
$\Osc_q \otimes \ldots \otimes \Osc_q = (\Osc_q)^{\otimes \, l}$, and denote
\begin{gather*}
b_i = 1 \otimes \ldots \otimes 1 \otimes b \otimes 1 \otimes \ldots \otimes 1, \qquad
b_i^\dagger = 1 \otimes \ldots \otimes 1 \otimes b^\dagger \otimes 1 \otimes \ldots \otimes 1, \\
q^{\nu N_i} = 1 \otimes \ldots \otimes 1 \otimes q^{\nu N} \otimes 1 \otimes \ldots \otimes 1, 
\end{gather*}
where $b$, $b^\dagger$ and $q^{\nu N}$ occupy only the $i$-th place of
the respective tensor products.

It was shown in \cite{NirRaz16b} that the mapping $\rho : \uqbp \to (\Osc_q)^{\otimes \, l}$ 
defined by the relations
\begin{align}
& \rho(q^{\nu h_0}) = q^{\nu (2 N_1 + \sum_{j = 2}^l N_j)}, 
&& \rho(e_0) = b^\dagger_1 \, q^{\sum_{j = 2}^l N_j}, 
\label{roh0e0}\\[.3em]
& \rho(q^{\nu h_i}) = q^{\nu (N_{i + 1} - N_{i})}, 
&& \rho(e_i) = - b^\ms_i \, b^{\mathstrut \dagger}_{i + 1} \, q^{N_i - N_{i + 1} - 1}, 
\label{rohiei}\\[.3em]
& \rho(q^{\nu h_l}) = q^{- \nu (2 N_l + \sum_{j = 1}^{l - 1} N_j)},
&& \rho(e_l) = - \kappa_q^{-1} \, b^\ms_l \, q^{N_l},
\label{rohlel}
\end{align}
where $i = 1, \ldots, l - 1$, is a homomorphism from the Borel subalgebra $\uqbp$
to the respective tensor power of the $q$-oscillator algebra. The homomorphism $\rho$ 
was obtained in the paper \cite{NirRaz16b}, starting with Jimbo's homomorphism from 
$\uqlsllpo$ to $\uqgllpo$, as the result of the interpretation of the defining relations 
for a subquotient of a degeneration of a shifted $\uqbp$-module. To get a representation 
of $\uqbp$, one chooses some representation of $(\Osc_q)^{\otimes \, l}$, and then takes 
the composition of this representation with $\rho$. 

It should be noted that a similar expression for $q$-oscillator representations was 
suggested by Kojima \cite{Koj08}. Our approach, however, gives such a benefit that 
we come to formulas (\ref{roh0e0})--(\ref{rohlel}) by degeneration of shifted Verma 
modules. With this, one can present $Q$-operators as a certain limit of transfer 
operators, see papers \cite{BooGoeKluNirRaz13, BooGoeKluNirRaz14a, NirRaz16a, BooGoeKluNirRaz14b} for the case $l = 1, 2$.

\section{\texorpdfstring{Automorphisms and more representations}{Automorphisms and
more representations}}

Other homomorphisms of type $\rho$ are obtained with the help of twists by  
automorphisms of $\uqbp$. Specifically to the case under consideration, there 
are two types of the automorphisms we use for such twisting. They originate 
from the automorphisms $\sigma$ and $\tau$ of the extended Dynkin diagram of 
$\sllpo$, where $\sigma$ acts by cyclic permutations of the roots 
$\sigma : \alpha_i \to \alpha_{i+1}$, while $\tau$ permutes the 
roots as $\tau : \alpha_i \to \alpha_{l + 1 - i}$, leaving 
$\alpha_0$ alone. Here, one has $\sigma^{l + 1} = \id$ and $\tau^2 = \id$. 
Applied to $\uqbp$, with a slight abuse of notation, their analogs produce 
the symmetry transformations
\begin{equation*}
\sigma(q^{\nu h_i}) = q^{\nu h_{i + 1}}, \qquad \sigma(e_i) = e_{i + 1}, \qquad
i \in \hi,
\end{equation*}
where it is assumed that $q^{\nu h_{l + 1}} = q^{\nu h_0}$ and $e_{l + 1} = e_0$, 
and
\begin{equation*}
\tau(q^{h_0}) = q^{h_0}, \qquad \tau(q^{h_i}) = q^{h_{l - i + 1}},
\qquad
\tau(e_0) = e_0, \qquad \tau(e_i) = e_{l - i + 1},
\qquad i \in I. 
\label{taueh}
\end{equation*}
It is clear that $\sigma^{l + 1}$ and $\tau^2$ are the identity transformations. 

Now we define
\begin{equation}
\rho_a = \rho \circ \sigma^{-a}, \qquad 
\overline\rho_a = \rho \circ \tau \circ \sigma^{-a+1}, 
\qquad a = 1, \ldots, l+1.
\label{roaovroa}
\end{equation}
It is worthwhile noting that $\ovrho_a$ in (\ref{roaovroa}) can be rewritten 
also using the relation
\begin{equation*}
\tau \circ \sigma^{-a + 1} = \sigma^{a - 1} \circ \tau 
= \sigma^{a - l - 2} \circ \tau, \qquad a = 1, \ldots, l+1.
\end{equation*}
Having the above definition, we obtain from (\ref{roh0e0})--(\ref{rohlel})
\begin{align}
& \rho_a(q^{\nu h_i}) = q^{\nu (N_{i - a +1} - N_{i - a})}, 
&& i = a + 1, \ldots, l, \ldots, l + a -1, \label{rohix} \\[.3em]
& \rho_a(q^{\nu h_{a - 1}}) = q^{- \nu (2 N_l + \sum_{j=1}^{l-1} N_j)}, 
 && \rho_a(q^{\nu h_{a}}) = q^{\nu (2 N_1 + \sum_{j=2}^l N_j)}, 
\label{rohax} \\[.3em] 
& \rho_a(e_i) = - b^{}_{i - a} \, b^\dagger_{i - a + 1} \, 
q^{N_{i - a} - N_{i - a + 1} - 1},
 && i = a + 1, \ldots, l, \ldots, l + a -1,  \label{roeix} \\[.3em]
& \rho_a(e_{a - 1}) = - \kappa_q^{-1} \, b^{}_{l} \, q^{N_{l}},   
&& \rho_a(e_a) = b^{\dagger}_{1} \, q^{\sum_{j=2}^l N_{j}}, \label{roeax}
\end{align} 
where $a = 1, \ldots, l+1$, and it is assumed that the index $i$ at the left 
hand side of (\ref{rohix}) and (\ref{roeix}) takes its values modulo $l + 1$, 
which means the identification $q^{\nu h_{l+1}} = q^{\nu h_{0}}$ and $e_{l+1} = e_{0}$. 

Further, applying certain tensor products of the representations $\chi^-$ and 
$\chi^+$, we define the homomorphisms $\theta_a$ as
\begin{equation}
\theta_a = (\underbracket[.6pt]{\chi^- \otimes \cdots \otimes \chi^-}_{l - a + 1} \otimes 
\underbracket[.6pt]{\chi^+ \otimes \cdots \otimes \chi^+}_{a - 1}) \circ \rho_a, \qquad 
a = 1, \ldots, l+1.
\label{thetaa}
\end{equation}
These representations are chosen so as to obtain highest $\ell$-weight representations. 
The corresponding basis vectors can be defined as
\begin{equation*}
v^{(a)}_{\bm m} = b_1^{m_1} \cdots b_{l - a + 1}^{m_{l-a+1}} \, 
b^{\dagger \, m_{l-a+2}}_{l - a + 2} \cdots b^{\dagger \, m_l}_{l} \, 
v_{\bf0},
\label{vam}
\end{equation*}
where $m_i \in \bbZ_+$ for all $i = 1, \ldots, l$ and we use the notation 
$\bm m = (m_1, \ldots, m_l)$ and $v_{\bf0} = v_{(0, \, \ldots \, , \, 0)}$.

For the mappings $\ovrho_a$, $a = 1, \ldots, l+1$, we obtain the following 
relations:
\begin{align*}
& \ovrho_a(q^{\nu h_i}) = q^{\nu (N_{a - i} - N_{a - i - 1})}, 
&& i = 0, 1, \ldots, a - 2,  \\[.3em]
& \ovrho_a(q^{\nu h_{a - 1}}) = q^{\nu (2 N_1 + \sum_{j=2}^{l} N_j)}, 
 && \ovrho_a(q^{\nu h_{a}}) = q^{- \nu (2 N_l + \sum_{j=1}^{l-1} N_j)}, 
 \\[.3em] 
& \ovrho_a(q^{\nu h_i}) = q^{\nu (N_{l + a - i + 1} - N_{l + a - i})}, 
&& i = a + 1, a + 2, \ldots, l,  \\[.3em]
& \ovrho_a(e_i) = - b^{}_{a - i - 1} \, b^\dagger_{a - i} \, 
q^{N_{a - i - 1} - N_{a - i} - 1},
 && i = 0, 1, \ldots, a - 2,  \\[.3em]
& \ovrho_a(e_{a - 1}) =  b^{\dagger}_{1} \, q^{\sum_{j=2}^l N_{j}},
&& \ovrho_a(e_a) =  - \kappa_q^{-1} \, b^{}_{l} \, q^{N_{l}}, 
 \\[.3em]
& \ovrho_a(e_i) = - b^{}_{l + a - i} \, b^\dagger_{l + a - i + 1} \, 
q^{N_{l + a - i} - N_{l + a - i + 1} - 1},
 && i = a + 1, a + 2, \ldots, l. 
\end{align*} 
Respectively, the homomorphisms $\ovtheta_a$ allowing one to obtain highest 
$\ell$-weight representations are now defined as
\begin{equation*}
\ovtheta_a = (\underbracket[.6pt]{\chi^- \otimes \cdots \otimes \chi^-}_{a - 1} \otimes 
\underbracket[.6pt]{\chi^+ \otimes \cdots \otimes \chi^+}_{l - a + 1}) \circ \ovrho_a, 
\qquad a = 1, \ldots, l+1.
\label{ovthetaa}
\end{equation*}
Then the corresponding basis vectors are given by
\begin{equation*}
\ovv^{(a)}_{\bm m} = b_1^{m_1} \cdots b_{a - 1}^{m_{a - 1}} \, b^{\dagger \, m_a}_{a \phantom{1}} 
\cdots b^{\dagger \, m_l}_l \, v_{\bm 0}.
\label{ovvam}
\end{equation*}
It appears that the vectors $v^{(a)}_{\bm m}$ and $\ovv^{(a)}_{\bm m}$ are $\ell$-weight vectors for 
the corresponding representations $\theta_a$ and $\ovtheta_a$, see below.

Inspired by the theory of quantum integrable systems, we associate with a representation of the quantum loop algebra 
a family of representations parametrized by the so-called spectral parameter. The usual way to do this is as follows. 
Given $\zeta \in \bbC^\times$, we define an automorphism $\Gamma_\zeta$ of $\uqlsllpo$ by its action on the generators 
as
\begin{equation*}
\Gamma_\zeta(e_i) = \zeta^{s_i} e_i, \qquad 
\Gamma_\zeta(f_i) = \zeta^{-s_i} f_i, \qquad 
\Gamma_\zeta(q^x) = q^x,
\end{equation*}
where $s_i$ are arbitrary integers. Note that the Borel subalgebras $\uqbp$ and $\uqbm$ are invariant with respect to the action of this automorphism. Now, if $\varphi$ is a representation of $\uqlsllpo$ or its Borel subalgebra, we define
\begin{equation*}
\varphi_\zeta = \varphi \circ \Gamma_\zeta.
\end{equation*}
If $V$ is the module corresponding to the representation $\varphi$, we denote by $V_\zeta$ 
the module corresponding to the representation $\varphi_\zeta$.
  
Note that for a representation $\varphi$ of $\uqlsllpo$ we have 
\begin{equation}
\varphi_\zeta(\phi^+_i(u)) = \varphi(\phi^+_i(\zeta^s u)), \qquad \varphi_\zeta(\phi^-_i(u^{-1})) = \varphi(\phi^-_i(\zeta^{-s} u^{-1})), \label{vpiu}
\end{equation}
where $s = s_0 + s_1 + \cdots + s_l$. If $\varphi$ is a representation of a Borel subalgebra of the quantum loop algebra $\uqlsllpo$, we have only one of the above equations.

In our case, starting from the representations $\theta_a$ and $\ovtheta_a$ we define the families $(\theta_a)_\zeta$ and $(\ovtheta_a)_\zeta$ as
\begin{equation*}
(\theta_a)_\zeta = \theta_a \circ \Gamma_\zeta, \qquad (\ovtheta_a)_\zeta = \ovtheta_a \circ \Gamma_\zeta.
\end{equation*}
The vectors $v^{(a)}_{\bm m}$ and $\ovv^{(a)}_{\bm m}$ are $\ell$-weight vectors for the representations 
$(\theta_a)_\zeta$ and $(\ovtheta_a)_\zeta$ as well. We use for the corresponding $\ell$-weights the notation 
determined by the equations
\begin{align*}
& (\theta_a)_\zeta (\phi^+_i(u)) \, v^{(a)}_{\bm m} = \theta_a(\phi^+_i(\zeta^s u)) \, v^{(a)}_{\bm m} 
= \Psi^+_{i, \, \bm m, \, a}(u) \, v^{(a)}_{\bm m}, \\
& (\ovtheta_a)_\zeta (\phi^+_i(u)) \, \ovv^{(a)}_{\bm m} = \ovtheta_a (\phi^+_i(\zeta^s u)) \, \ovv^{(a)}_{\bm m} 
= \ovPsi^+_{i, \, \bm m, \, a}(u) \, \ovv^{(a)}_{\bm m},
\end{align*}
where we take into account the first equation of (\ref{vpiu}). The corresponding elements of $\gothh^*$ are denoted 
as $\lambda_{\bm m, \, a}$ and $\ovlambda_{\bm m, \, a}$.

\subsection{\texorpdfstring{Representations $\theta_a$}{Representations thetaa}} 
\label{ss:a}

\subsubsection{\texorpdfstring{$\theta_a$ : $a = 1$}{a = 1}} \label{ss:a1}

By putting $a = 1$ in (\ref{rohix})--(\ref{roeax}), we come to the homomorphism
$\rho_1$ given by the following relations:
\begin{align*}
& \rho_1(q^{\nu h_{0}}) = q^{- \nu (2 N_l + \sum_{j=1}^{l-1} N_j)}, 
&& \rho_1(q^{\nu h_{1}}) = q^{\nu (2 N_1 + \sum_{j=2}^l N_j)},  \\[.3em]
& \rho_1(q^{\nu h_i}) = q^{\nu (N_{i} - N_{i - 1})}, 
&& i = 2, \ldots, l,  \\[.3em]
& \rho_1(e_{0}) = - \kappa_q^{-1} \, b^{}_{l} \, q^{N_{l}},   
&& \rho_1(e_1) = b^{\dagger}_{1} \, q^{\sum_{j=2}^l N_{j}}, \\[.3em]
& \rho_1(e_i) = - b^{}_{i - 1} \, b^\dagger_{i} \, 
q^{N_{i - 1} - N_{i} - 1}, && i = 2, \ldots, l.
\end{align*}
It is easy to see that
\begin{equation}
\lambda_{\bm m, \, 1} = - \Big( 2 m_1 + \sum_{j = 2}^l m_j + l + 1 \Big) \, \omega_1 
- \sum_{i = 2}^l (m_i - m_{i - 1}) \, \omega_i. \label{lambdam1}
\end{equation}
Here and below $\omega_i \in \gothh^*$, $i \in I$, are the fundamental weights defined by the relations
\begin{equation*}
\langle \omega_i, \, h_j \rangle = \delta_{i j}, \qquad j \in I.
\end{equation*}

Using equation (\ref{edma1}), we calculate
\begin{equation*}
\rho_1(e_{\delta - \alpha_1}) = - \kappa_q^{-1} \, b^{}_1 \, 
q^{\sum^l_{j=1} N_j + l - 1}
\label{ro1edma1}
\end{equation*}
and therefore, from (\ref{epd}) for $\gamma = \alpha_1$, we obtain
\begin{equation*}
\rho_1(e'_{\delta, \, \alpha_1}) = - \kappa_q^{-1} \big(1 - [2]_q \, q^{2 N_1 + 1} \big)
q^{2 \sum^l_{j = 2} N_j + l}.
\label{ro1eda1}
\end{equation*}
Further, from the first relation of (\ref{cwbx}) for $\gamma = \alpha_1$ 
we subsequently derive
\begin{equation*}
\rho_1(e_{\alpha_1 + n\delta}) = (-1)^n \, b^\dagger_1 \, 
q^{2n N_1 + (2n+1)\sum_{j=2}^l N_j + 2 n + n l}, \qquad n \in \bbZ_+,
\label{ro1eand}
\end{equation*}
and so, the first relation from (\ref{cwbz}) implies
\begin{equation*}
\rho_1(e'_{n\delta, \, \alpha_1}) = \kappa_q^{-1} \, (-1)^n \big( [n]_q \, q^{-2N_1 - 1}
- [n+1]_q \big) q^{n(2\sum_{j=1}^l N_j + l + 1)}, \qquad n \in \bbN.
\label{ro1epnda1}
\end{equation*}
Then, seeing from (\ref{oi}) that $o_1 = 1$, we obtain from (\ref{phipiu}) the
relation
\begin{multline}
\rho_1(\phi_1^+(u)) = q^{2N_1 + \sum_{j=2}^l N_j} \,
\big(1 - q^{2\sum_{j=2}^l N_j + l} \, u \big) \\* 
\times \big( 1 - q^{2\sum_{j=1}^l N_j + l} \, u \big)^{-1} \, 
\big( 1 - q^{2\sum_{j=1}^l N_j + l + 2} \, u \big)^{-1}.
\label{ro1phi1u}
\end{multline}
Now, composing $\rho_1$ with $\chi^-$ up to the representation $\theta_1$ 
just according to (\ref{thetaa}), we obtain from (\ref{ro1phi1u}) the 
following equation:
\begin{equation*}
\phi_1^+(u) \, v^{(1)}_{\bm m} = \frac{q^{-2m_1 - \sum_{j=2}^l m_j - l - 1} 
(1 - q^{-2\sum_{j=2}^l m_j - l + 2} \, u)} {(1 - q^{-2\sum_{j=1}^l m_j - l} \, u)
(1 - q^{-2\sum_{j=1}^l m_j - l + 2} \, u)} \, v^{(1)}_{\bm m}.
\label{phi1uv1m}
\end{equation*}

Further, from (\ref{edmai})--(\ref{edmal}) we obtain
\begin{equation*}
\rho_1(e_{\delta - \alpha_i}) = (-1)^i \, b^{}_{i} \, b^\dagger_{i-1} \,
q^{2 \sum_{j = i}^l N_j + l - i}, \qquad i = 2, \ldots, l,
\label{ro1edmai}
\end{equation*}
and consequently
\begin{equation*}
\rho_1(e'_{\delta, \, \alpha_i}) = (-1)^i \, \kappa_q^{-1} \, 
\big(1 - [2]_q \, q^{2N_i + 1} + q^{2N_{i-1} + 2N_i + 2} \big) \, 
q^{2 \sum_{j = i+1}^l N_j + l - i + 1}.
\label{epdai}
\end{equation*}
Using again the first relation from (\ref{cwbx}), we subsequently derive
the equation
\begin{equation*}
\rho_1(e_{\alpha_i + n\delta}) = - (-1)^{ni} \, b^{}_{i-1} \, b^\dagger_i \,
q^{N_{i-1} - N_i - 1 + 2n \sum_{j=i}^l N_j + n(l - i + 3)}, \qquad n \in \bbZ_+,
\label{ro1eaind}
\end{equation*}
and hence,
\begin{multline*}
\rho_1(e'_{n\delta, \, \alpha_i}) = - (-1)^{ni} \, \kappa_q^{-1} \biggl( [n-1]_q \, 
q^{2N_{i-1} - 2N_i} \\*
- [n]_q \big( q^{2N_{i-1} + 1} + q^{-2N_i - 1} \big)
+ [n+1]_q \biggr) q^{2n \sum_{j = i}^l N_j + n(l - i + 2)}, \qquad n \in \bbN.
\label{ro1epndai}
\end{multline*}
As a consequence, relation (\ref{phipiu}) gives for $\phi^+_i(u)$, 
$i = 2, \ldots, l$, the following expression:
\begin{multline*}
\rho_1(\phi_i^+(u)) = q^{N_i - N_{i-1}} \, 
\big( 1 - q^{2\sum_{j=i-1}^l N_j + l - i + 3} \, u \big) \,
\big( 1 - q^{2\sum_{j=i+1}^l N_j + l - i + 1} \, u \big) \\* 
\times \big( 1 - q^{2\sum_{j=i}^l N_j + l - i + 3} \, u \big)^{-1} \,
\big( 1 - q^{2\sum_{j=i}^l N_j + l - i + 1} \, u \big)^{-1}. 
\end{multline*}
Therefore, referring again to the representation $\theta_1$ defined by (\ref{thetaa}) 
with $a = 1$, we obtain
\begin{equation*}
\phi_i^+(u) \, v^{(1)}_{\bm m} = q^{m_{i-1} - m_i} \, 
\frac{(1 - q^{-2\sum_{j=i-1}^l m_j - l + i - 1} \, u)
(1 - q^{-2\sum_{j=i+1}^l m_j - l + i + 1} \, u)}
{(1 - q^{-2\sum_{j=i}^l m_j - l + i - 1} \, u)
(1 - q^{-2\sum_{j=i}^l m_j - l + i + 1} \, u)} \, v^{(1)}_{\bm m},
\label{phiiuv1m}
\end{equation*}
where $i = 2, \ldots, l$. We have the explicit expressions
\begin{align*}
\Psi^+_{1, \, {\bm m}, \, 1}(u) & = \frac{q^{-2m_1 - \sum_{j=2}^l m_j - l - 1} 
(1 - q^{-2\sum_{j=2}^l m_j - l + 2} \, \zeta^s \, u)} {(1 - q^{-2\sum_{j=1}^l m_j - l} \, 
\zeta^s \, u) (1 - q^{-2\sum_{j=1}^l m_j - l + 2} \, \zeta^s \, u)}, \\
\intertext{and}
\Psi^+_{i, \, {\bm m}, \, 1}(u) & = q^{m_{i-1} - m_i} \, 
\frac{(1 - q^{-2\sum_{j=i-1}^l m_j - l + i - 1} \, \zeta^s \, u)
(1 - q^{-2\sum_{j=i+1}^l m_j - l + i + 1} \, \zeta^s \, u)}
{(1 - q^{-2\sum_{j=i}^l m_j - l + i - 1} \, \zeta^s \, u)
(1 - q^{-2\sum_{j=i}^l m_j - l + i + 1} \, \zeta^s \, u)} 
\end{align*}
for $i = 2, \ldots, l$. Together with (\ref{lambdam1}), they give
\begin{equation*}
\lambda_{\bm 0, \, 1} = - (l + 1) \, \omega_1, \qquad 
\Psi^+_{1, \, {\bf0}, \, 1}(u) = q^{- l - 1}(1 - q^{- l} \, \zeta^s \, u)^{-1}, 
\qquad
\Psi^+_{i, \, {\bf0}, \, 1}(u) = 1, 
\end{equation*}
where $i = 2, \ldots, l$.  
We have thus established that the representation $(\theta_1)_\zeta$ is isomorphic 
to a shifted prefundamental representation of $\uqbp$.

\subsubsection{\texorpdfstring{$\theta_a$ : $a = 2, \ldots, l$}{a = 2,...,l}} 
\label{ss:gen-a}

In this case, as follows from (\ref{rohix})--(\ref{rohax}) and 
(\ref{roeix})--(\ref{roeax}), we have the following relations defining 
the homomorphisms $\rho_a$ for $a$ with values from $2$ to $l$:
\begin{align}
& \rho_a(q^{\nu h_i}) = q^{\nu (N_{l + i - a + 2} - N_{l + i - a + 1})},
&& i = 0,1,\ldots,a-2, \label{roahix2} \\
& \rho_a(q^{\nu h_{a-1}}) = q^{-\nu (2N_l + \sum_{j=1}^{l - 1} N_j)}, 
&& \rho_a(q^{\nu h_a}) = q^{\nu (2N_1 + \sum_{j=2}^l N_j)}, \label{roahax2} \\
& \rho_a(q^{\nu h_i}) = q^{\nu (N_{i - a + 1} - N_{i - a})},
&& i = a + 1, \ldots, l, \label{roahixx2} \\
& \rho_a(e_i) = - b^{}_{l + i - a + 1} \, b^\dagger_{l + i - a + 2} \,
q^{N_{l + i - a + 1} - N_{l + i - a + 2} - 1}, && i = 0,1,\ldots,a-2, 
\label{roaeix2} \\
& \rho_a(e_{a-1}) = - \kappa_q^{-1} \, b^{}_l \, q^{N_l}, &&
\rho_a(e_a) = b^\dagger_1 \, q^{\sum_{j=2}^l N_j}, \label{roaeax2} \\
& \rho_a(e_i) = - b^{}_{i - a} \, b^\dagger_{i - a + 1} \,
q^{N_{i - a} - N_{i - a + 1} - 1}, && i = a + 1, \ldots, l. \label{roaeixx2}  
\end{align}
We determine that in this case
\begin{multline}
\lambda_{\bm m, \, a} = \sum_{i = 1}^{a - 2} (m_{l + i - a + 2} - m_{l + i - a + 1}) \, \omega_i 
- \sum_{i = a + 1}^l (m_{i - a + 1} - m_{i - a}) \, \omega_i \\
+ \Big( \sum_{j = 1}^{l - a + 1} m_j - \sum_{j = l - a + 2}^{l - 1} m_j - 2 m_l + l - a + 1 \Big) \, 
\omega_{a - 1} \\
- \Big( 2 m_1 + \sum_{j = 2}^{l - a + 1} m_j - \sum_{j = l - a + 2}^l m_j + l - a + 2 \Big) \, \omega_a.
\label{lambdama}
\end{multline}

To find expressions for the functions $\Psi^+_{i, \, {\bm m}, \, a}(u)$ we first obtain some subsidiary 
relations based on equations (\ref{roaeix2}) and (\ref{roaeixx2}). These are
\begin{multline}
\rho_a(e_{\delta - \alpha_{1, \, l - k}}) = 
\rho_a([e_{\alpha_{l-k}} \, , \ldots [e_{\alpha_l} \, , \, e_{\delta-\theta}]_q \ldots ]_q) 
\\* = - b^{}_{l - a - k} \, b^\dagger_{l - a + 2} \,
q^{\sum_{j=l-a-k}^{l-a+1} N_{j} - N_{l - a + 2} + k} \label{subfor}
\end{multline}
for $0 \le k \le l - a - 1$. Putting $k = l - a - 1$ and continuing with $e_{\alpha_a}$
from (\ref{roaeax2}), we obtain 
\begin{multline*}
\rho_a(e_{\delta - \alpha_{1 a}}) = \rho_a([e_{\alpha_{a}} \, , \, [e_{\alpha_{a+1}} \, , 
\ldots [e_{\alpha_l} \, , \, e_{\delta-\theta}]_q \ldots ]_q ]_q) \\* 
= b^\dagger_{l - a + 2} \, q^{\sum_{j = 1}^{l} N_j + \sum_{j = 1}^{l - a + 1} N_j 
- N_{l - a + 2} + l - a + 1}, \qquad a = 2, \ldots, l.
\end{multline*}
Then it follows from this relation and expression (\ref{roaeax2}) for the image of
$e_{\alpha_{a-1}}$ immediately that
\begin{equation*}
\rho_a(e_{\delta - \alpha_{1, \, a - 1}}) = 0
\end{equation*}
and therefore,
\begin{equation*}
\rho_a(e_{\delta - \alpha_i}) = 0, \qquad \rho_a(e'_{\delta, \, \alpha_i}) = 0, 
\qquad i = 1, \ldots, a - 2, \quad a = 2, \ldots, l. 
\label{roaedmai}
\end{equation*}
As a consequence, we also have
\begin{equation*}
\rho_a(e'_{n\delta, \, \alpha_i}) = 0, \qquad i = 1, \ldots, a - 2, 
\quad a = 2, \ldots, l, 
\end{equation*}
We thus obtain that the image of $\phi^+_i(u)$ under the homomorphism
$\rho_a$ is fully determined by that of $q^{h_i}$, that is
\begin{equation*}
\rho_a(\phi_i^+) = q^{N_{l + i - a + 2} - N_{l + i - a + 1}}, 
\qquad i = 1, \ldots, a - 2, \quad a = 2, \ldots, l,
\label{roaphiiu}
\end{equation*}
and, for the given range of $i$ and $a$,
\begin{equation}
\Psi^+_{i, \, {\bm m}, \, a}(u) = q^{m_{l + i - a + 2} - m_{l + i - a + 1}}.
\label{8.67}
\end{equation}

Now we consider two special cases, $i = a - 1$ and $i = a$. For the first case we obtain
\begin{equation*}
\rho_a(e_{\delta - \alpha_{a-1}}) = (-1)^a \, b^\dagger_l \, 
q^{2 \sum_{j=1}^{l-a+1} N_j + l - a + 1},
\label{8.69}
\end{equation*}
and so,
\begin{equation*}
\rho_a(e'_{\delta, \, \alpha_{a-1}}) = -(-1)^a \, \kappa_q^{-1} \, 
q^{2\sum_{j=1}^{l-a+1} N_j + l - a + 2}.
\label{8.70}
\end{equation*}
Implementing the procedure expressed by (\ref{cwbx})--(\ref{cwbz}), we obtain that the images 
under the homomorphism $\rho_a$ of all higher root vectors $e_{\alpha_{a-1} + n\delta}$ for 
$n > 0$ vanish, and we thus have
\begin{equation}
\rho_a(e_{\alpha_{a-1} + n\delta}) = \delta_{n,0} \, 
\big(-\kappa_q^{-1} \, b^{}_l \, q^{N_l} \big), \qquad n \in \bbZ_+.
\label{8.71}
\end{equation}
Then it follows from (\ref{8.71}) that only $e'_{\delta, \, \alpha_{a-1}}$ survives among
all $e'_{n\delta, \, \alpha_{a-1}}$, and so, we have
\begin{equation*}
\rho_a(e'_{n\delta, \, \alpha_{a-1}}) = \delta_{n,1} \, \big(-(-1)^a \, \kappa_q^{-1} \, 
q^{2\sum_{j=1}^{l-a+1} N_j + l - a + 2} \big), \qquad n \in \bbN. 
\label{8.72}
\end{equation*}
Using relation (\ref{phipiu}) we obtain
\begin{equation*}
\rho_a(\phi^+_{a-1}(u)) = q^{-2N_l - \sum_{j=1}^{l-1} N_j} \, 
\big( 1 - q^{2\sum_{j=1}^{l-a+1} N_j + l - a + 2} \, u \big).
\label{8.73}
\end{equation*}
Applying the map $\theta_a$ according to (\ref{thetaa}) for $a = 2, \ldots, l$, 
we arrive at the formula
\begin{equation}
\Psi^+_{a-1, \, {\bm m}, \, a}(u) = q^{\sum_{j=1}^{l-a+1} m_j 
- \sum_{j = l - a + 2}^{l-1} m_j - 2 m_l + l - a + 1} \, 
\big( 1 - q^{-2 \sum_{j=1}^{l - a + 1} m_j - l + a} \, \zeta^s \, u \big).
\label{8.74}
\end{equation}

For the second special case, $i=a$, using subsidiary formula (\ref{subfor}) for 
$k = l - a - 1$, we first obtain another useful relation
\begin{multline*}
\rho_a(e_{\delta - \alpha_{k + 1, \, a + 1}}) = \rho_a([e_{\alpha_{k}} \, , \ldots 
[e_{1} \, , \, [e_{\alpha_{a + 1}} 
\, , \ldots [e_{\alpha_l} \, , \, e_{\delta-\theta}]_q \ldots 
]_q ]_q \ldots ]_q) \\ =
- (-1)^k \, b^{}_1 \, b^{\dagger}_{l - a + 2 + k} \, q^{\sum_{j = 1}^{l - a + 1} N_j 
- \sum_{j = l - a + 2}^{l - a + 2 + k} N_j + l - a - 1}, 
\end{multline*}
where the integer $k$ can take any value from $1$ to $a - 2$. Putting here
$k = a - 2$ and using the image of $e_{\alpha_{a-1}}$ under $\rho_a$ from
(\ref{roaeax2}), we obtain
\begin{equation*}
\rho_a(e_{\delta - \alpha_a}) = (-1)^a \, \kappa_q^{-1} \, b^{}_1 \, 
q^{\sum_{j=1}^{l - a + 1} N_j - \sum_{j = l - a + 2}^{l} N_j + l - a}.
\label{8.75}
\end{equation*}
Then we obtain
\begin{equation*}
\rho_a(e'_{\delta, \, \alpha_a}) = (-1)^a \, \kappa_q^{-1} \, 
\big( 1 - [2]_q \, q^{2N_1 + 1} \big)\, q^{2\sum_{j=2}^{l - a + 1} N_j + l - a + 1}.
\label{8.76}
\end{equation*}
According to (\ref{cwbx})--(\ref{cwbz}), it gives for the image of the 
higher root vectors the expression 
\begin{equation*}
\rho_a(e_{\alpha_a + n\delta}) = (-1)^{n a} \, b^\dagger_1 \, 
q^{\sum_{j=2}^l N_j + 2n \sum_{j=1}^{l-a+1} N_j + n l - n a + 3 n}, \qquad
n \in \bbZ_+, \label{8.77}
\end{equation*}
and we obtain
\begin{equation*}
\rho_a(e'_{n\delta, \, \alpha_a}) = (-1)^{na} \, \kappa_q^{-1} \, 
\big( [n]_q \, q^{-2N_1 - 1} - [n+1]_q \big) \, q^{2n \sum_{j=1}^{l-a+1} N_j +
n(l - a + 2)}, \qquad n \in \bbN.
\label{8.78}
\end{equation*}
Then, taking into account (\ref{phipiu}) and (\ref{roahax2}), we see that
for $\phi^+_a(u)$ the following equation holds:
\begin{multline*}
\rho_a(\phi_a^+(u)) = q^{2N_1 + \sum_{j=2}^l N_j} \, 
\big( 1 - q^{2 \sum_{j=2}^{l - a + 1} N_j + l - a + 1} \, u \big) \\
\times \big( 1 - q^{2 \sum_{j=1}^{l - a + 1} N_j + l - a + 1} \, u \big)^{-1} 
\big( 1 - q^{2 \sum_{j=1}^{l - a + 1} N_j + l - a + 3} \, u \big)^{-1}.
\label{8.79}
\end{multline*}
Therefore, in the representation $\theta_a$ defined by (\ref{thetaa}) we have
\begin{equation}
\Psi^{+}_{a, \, {\bm m}, \, a}(u) = \frac{q^{-2m_1 - \sum_{j=2}^{l-a+1} m_j +
\sum_{j = l - a + 2}^{l} m_j - l + a - 2} \, \big( 1 - q^{- 2 \sum_{j=2}^{l-a+1} m_j 
- l + a + 1} \, \zeta^s \, u \big) }{\big( 1 - q^{- 2 \sum_{j=1}^{l - a + 1} m_j - l + a - 1} 
\, \zeta^s \, u \big) \, \big( 1 - q^{- 2 \sum_{j=1}^{l - a + 1} m_j - l + a + 1} \, 
\zeta^s \, u \big)}.
\label{8.80}
\end{equation}

Consider now the case $i = a + 1, \ldots, l$. For this case we should basically use
relations (\ref{roahixx2}), (\ref{roaeixx2}). We first obtain
\begin{multline*}
\rho_a(e_{\delta - \alpha_{1, \, i + 1}}) = \rho_a \big( [e_{\alpha_{i + 1}} \, , \ldots [e_{\alpha_l} \, , \, e_{\delta-\theta}]_q 
\ldots ]_q \big) \\
= - b^{}_{i - a + 1} \, b^{\dagger}_{l-a+2} \, 
q^{\sum_{j=i-a+1}^{l-a+1} N_j - N_{l-a+2} + l - i - 1}.
\label{8.81x}
\end{multline*}
Then, continuing in accordance with (\ref{edmai}), we obtain the subsidiary relation
\begin{multline*}
\rho_a(e_{\delta - \alpha_{k + 1, \, i + 1}}) = \rho_a ( [e_{\alpha_k} \, , \, \ldots 
[e_{\alpha_1} \, , \, [e_{\alpha_{i + 1}} \, , \, \ldots 
[e_{\alpha_l} \, , \, e_{\delta - \theta}]_q \ldots ]_q ]_q \ldots ]_q ) \\ =
- (-1)^k \, b^{}_{i - a + 1} \, b^\dagger_{l - a + 2 + k} \,
q^{\sum_{j = i - a + 1}^{l - a + 1} N_j - \sum_{j = l - a + 2}^{l - a + 2 + k} N_j 
+ l - i - 1}
\end{multline*}
valid for $k = 1,\ldots,a-2$. Putting here $k = a - 2$ and continuing with
$e_{\alpha_{a-1}}$ and $e_{\alpha_a}$ from (\ref{roaeax2}) we come to the
formula
\begin{multline*}
\rho_a(e_{\delta - \alpha_{a + k + 1, \, i + 1}}) \\
= \rho_a ( [e_{\alpha_{a+k}} \, , \, \ldots [e_{\alpha_a} \, , \, [e_{\alpha_{a-1}} \, , \, \ldots 
[e_{\alpha_1} \, , \, [e_{\alpha_{i + 1}} \, , \, \ldots 
[e_{\alpha_l} \, , \, e_{\delta - \theta}]_q \ldots ]_q ]_q \ldots ]_q ]_q \ldots ]_q ) \\ 
= (-1)^{a+k+1} \, b^{}_{i - a + 1} \, b^\dagger_{k + 1} \,
q^{\sum_{j = i - a + 1}^{l - a + 1} N_j + \sum_{j = k + 2}^{l - a + 1} N_j 
+ l - i},
\end{multline*}
valid for $k = 0, 1, \ldots, i - a - 1$. Putting here $k = i - a - 1$, we obtain
\begin{equation*}
\rho_a(e_{\delta - \alpha_i}) = (-1)^i \, b^{}_{i-a+1} \, b^\dagger_{i-a} \,
q^{2\sum_{j=i-a+1}^{l - a + 1} N_j + l - i},
\label{8.81}
\end{equation*}
and consequently
\begin{equation*}
\rho_a(e'_{\delta, \, \alpha_i}) = (-1)^i \, \kappa_q^{-1} 
\big( 1 - [2]_q \, q^{2 N_{i-a+1} + 1} + q^{2N_{i-a} + 2N_{i-a+1} + 2} \big) 
q^{2\sum_{j=i-a+2}^{l-a+1} N_j + l - i + 1},
\label{8.82}
\end{equation*}
both relations valid for $i = a + 1, \ldots, l$ and $a = 2, \ldots, l$.
Respectively, we obtain
\begin{equation*}
\rho_a(e_{\alpha_i + n\delta}) = - (-1)^{n i} \, b^{}_{i-a} \, b^\dagger_{i-a+1} \,
q^{N_{i-a} - N_{i-a+1} - 1 + 2n \sum_{j=i-a+1}^{l-a+1} N_j + n(l - i + 3)}, \qquad
n \in \bbZ_+. 
\label{8.83}
\end{equation*}
The latter allows us to obtain
\begin{multline*}
\rho_a(e'_{n\delta, \, \alpha_i}) = - (-1)^{n i} \, \kappa_q^{-1} \, 
\biggl( [n-1]_q \, q^{2N_{i-a} - 2N_{i-a+1}} 
\\- [n]_q \, \big( q^{2N_{i-a} + 1} + q^{-2N_{i-a+1} - 1} \big) 
+ [n+1]_q \biggr) q^{2n \sum_{j=i-a+1}^{l-a+1} N_j + n(l - i + 2)}, 
\qquad n \in \bbN. 
\end{multline*}
Using the relation between $\phi^+_i(u)$ and $e'_{\delta, \, \alpha_i}(u)$ 
as given by (\ref{phipiu}), we obtain the following expression:
\begin{multline*}
\rho_a(\phi^+_i(u)) = q^{N_{i-a+1} - N_{i-a}} \, 
\big( 1 - q^{2\sum_{j=i-a+2}^{l-a+1} N_j + l - i + 1} \, u \big) 
\big( 1 - q^{2\sum_{j=i-a}^{l-a+1} N_j + l - i + 3} \, u \big) \\*
\times \big( 1 - q^{2\sum_{j=i-a+1}^{l-a+1} N_j + l - i + 1} \, u \big)^{-1}
\big( 1 - q^{2\sum_{j=i-a+1}^{l-a+1} N_j + l - i + 3} \, u \big)^{-1}.
\end{multline*}
Recalling our definition of the representations $\theta_a$ in (\ref{thetaa}), 
we finally obtain
\begin{multline}
\Psi^+_{i, \, {\bm m}, \, a}(u) \\
= q^{m_{i-a} - m_{i-a+1}} \, 
\frac{( 1 - q^{-2\sum_{j=i-a}^{l - a + 1} m_j - l + i - 1} \, \zeta^s \, u )
( 1 - q^{-2\sum_{j=i-a+2}^{l-a+1} m_j - l + i + 1} \, \zeta^s \, u )}
{( 1 - q^{-2\sum_{j=i-a+1}^{l - a + 1} m_j - l + i - 1} \, \zeta^s \, u )
( 1 - q^{-2\sum_{j=i-a+1}^{l - a + 1} m_j - l + i + 1} \, \zeta^s \, u )}
\label{8.86}
\end{multline}
for all $i = a + 1, \ldots, l$ and $a = 2,\ldots, l$. 
We thus see from (\ref{lambdama}), (\ref{8.67}), (\ref{8.74}), (\ref{8.80}) and (\ref{8.86}) that 
for $a = 2, \ldots, l$ the corresponding highest $\ell$-weights are determined 
by the expressions 
\begin{align*}
& \lambda_{\bm 0, \, a} = (l - a + 1) \, \omega_{a - 1} - (l - a + 2) \, \omega_a, \\
\intertext{and}
& \Psi^+_{i, \, {\bf0}, \, a}(u) = \left\{ 
\begin{array}{cl} 
1, & i = 1,\ldots,a-2, \\
q^{l - a + 1} \, ( 1 - q^{- l + a} \, \zeta^s \, u ), & i = a-1, \\
q^{- l + a - 2} \, ( 1 - q^{- l + a - 1} \, \zeta^s \, u )^{-1}, & i = a, \\
1, & i = a+1, \ldots, l.
\end{array}
\right.
\label{8.89}
\end{align*}
It means that the representations $(\theta_a)_\zeta$ for $a = 2, \ldots, l$ are
not prefundamental, but are isomorphic to subquotients of tensor 
products of two shifted prefundamental representations of $\uqbp$.

\subsubsection{\texorpdfstring{$\theta_a$ : $a = l + 1$}{a = l + 1}} \label{ss:l+1}

In this case we have $\rho_{l + 1} = \rho$, therefore
\begin{align*}
& \rho_{l+1}(q^{\nu h_0}) = q^{\nu (2 N_1 + \sum_{j = 2}^l N_j)}, 
&& \rho_{l+1}(e_0) = b^\dagger_1 \, q^{\sum_{j = 2}^l N_j}, \\[.3em]
& \rho_{l+1}(q^{\nu h_i}) = q^{\nu (N_{i + 1} - N_{i})}, 
&& \rho_{l+1}(e_i) = - b^\ms_i \, b^{\mathstrut \dagger}_{i + 1} 
\, q^{N_i - N_{i + 1} - 1}, \\[.3em]
& \rho_{l+1}(q^{\nu h_l}) = q^{- \nu (2 N_l + \sum_{j = 1}^{l - 1} N_j)},
&& \rho_{l+1}(e_l) = - \kappa_q^{-1} \, b^\ms_l \, q^{N_l},
\end{align*}
where $i = 1, \ldots, l - 1$. Now we have
\begin{equation}
\lambda_{\bm m, \, l + 1} = \sum_{i = 1}^{l - 1} (m_{i + 1} - m_i) \, \omega_i 
- \Big( \sum_{j = 1}^{l - 1} m_j + 2 m_l \Big) \, \omega_l. \label{lambdamlpo}
\end{equation}

Similarly to the preceding, we obtain
\begin{equation*}
\rho_{l+1}(e_{\delta - \alpha_i}) = 0, \quad i = 1, \ldots, l - 1, \qquad 
\rho_{l+1}(e_{\delta - \alpha_l}) = - (-1)^l \, b^\dagger_l,
\label{8.7x}
\end{equation*}
so that,
\begin{equation*}
\rho_{l+1}(e'_{\delta, \, \alpha_i}) = 0, \quad i = 1, \ldots, l - 1, \qquad 
\rho_{l+1}(e'_{\delta, \, \alpha_l}) = (-1)^l \, \kappa_q^{-1} \, q.
\label{8.9x}
\end{equation*}  
Therefore, we obtain
\begin{equation*}
\rho_{l+1}(1 - \kappa_q \, e'_{\delta, \, \alpha_i}(u)) = 1, \quad i = 1, \ldots, l - 1, 
\qquad \rho_{l+1}(1 - \kappa_q \, e'_{\delta, \alpha_l}(u)) = 1 - (-1)^l \, q \, u.
\label{8.12x}
\end{equation*}
This allows us to write down the equations
\begin{equation*}
\rho_{l+1}(\phi^+_i(u)) = \rho_{l+1}(q^{h_i}) = q^{N_{i+1} - N_i}, \qquad
i = 1, \ldots, l - 1, 
\end{equation*}
and
\begin{equation*}
\rho_{l+1}(\phi^+_l(u)) = q^{- 2 N_l - \sum_{j = 1}^{l-1} N_j} \, (1 - q \, u)
\end{equation*}
for the images of $\phi^+_i$ under $\rho_{l+1}$. Then, using the representation 
$\theta_{l+1}$ we see that 
\begin{equation*}
\Psi^+_{i, \, {\bm m}, \, l+1}(u) = q^{m_{i+1} - m_i}, \quad 
i = 1, \ldots, l - 1, \qquad
\Psi^+_{l, \, {\bm m}, \, l+1}(u) = q^{-2m_l - \sum_{j=1}^{l-1} m_j} \,
(1 - q \, \zeta^s \, u).
\end{equation*}
We obtain from (\ref{lambdamlpo}) and from the last line that 
\begin{equation*}
\lambda_{\bm 0, \, l + 1} = 0, \qquad \Psi^+_{i, \, {\bf0}, \, l + 1}(u) = 1, \qquad 
i = 1, \ldots, l - 1, \qquad
\Psi^+_{l, \, {\bf0}, \, l + 1}(u) = (1 - q \, \zeta^s \, u).
\end{equation*}
Therefore, we see that the representation $(\theta_{l+1})_\zeta$ is isomorphic to a 
prefundamental representation.

\subsection{\texorpdfstring{Representations $\ovtheta_a$}{Representations ovthetaa}} \label{ss:abar}

For the representations $\ovtheta_a$ we first make the following observation:
\begin{equation*}
\ovrho_a(q^{\nu h_i}) = \rho_{l - a + 2}(q^{\nu h_{l - i + 1}}), 
\qquad \ovrho_a(e_i) = \rho_{l - a + 2}(e_{l - i + 1})
\label{8.b1}
\end{equation*}
and
\begin{equation*}
\ovv^{(a)}_{\bm m} = v^{(l - a + 2)}_{\bm m}, \qquad a = 1, \ldots, l + 1.
\end{equation*}
Applied to the relation between $\phi^+_i(u)$ and $e'_{n \delta, \, \alpha_i}$ in 
(\ref{phipiu}), this leads us immediately to the conclusion that the functions 
$\ovPsi^+_{i, \, {\bm m}, \, a}(u)$ are related to the functions 
$\Psi^+_{i, \, {\bm m}, \, a}(u)$ as
\begin{equation*}
\ovPsi^+_{i, \, {\bm m}, \, a}(u) 
= \Psi^+_{l - i + 1, \, {\bm m}, \, l - a + 2}(-(-1)^{l} \, u).
\label{8.b2}
\end{equation*}
For the elements $\ovlambda_{\bm m, \, a}$ we have
\begin{equation*}
\ovlambda_{\bm m, \, a} = \iota(\lambda_{\bm m, \, l - a + 2}),
\end{equation*}
where $\iota$ is a linear mapping determined by the relation
\begin{equation*}
\iota(\omega_i) = \omega_{l - i + 1}.
\end{equation*}
The above relations show that, similarly as in the case of the representations $(\theta_a)_\zeta$,  
the representation $(\ovtheta_1)_\zeta$ is isomorphic to a prefundamental representation, the 
representation $(\ovtheta_{l+1})_\zeta$ is isomorphic to a shifted prefundamental representation, 
while the others, $(\ovtheta_a)_\zeta$ for $a = 2, \ldots, l$, can be realized as 
subquotients of tensor products of two prefundamental representations of 
$\uqbp$.

\section{Conclusion}

We have obtained the $\ell$-weights and the corresponding $\ell$-weight vectors for 
the $q$-oscillator representations of the Borel subalgebra $\uqbp$ of the quantum 
loop algebra $\uqlsllpo$ with arbitrary rank $l$. Here we have discovered 
that for all representations $(\theta_a)_\zeta$, $(\ovtheta_a)_\zeta$ the 
representation space has a basis consisting of $\ell$-weight vectors. This 
means that the number $p$ in (\ref{intp}) is always equal to $1$. We see 
that some $q$-oscillator representations, namely $(\theta_1)_\zeta$, 
$(\theta_{l + 1})_\zeta$ and $(\ovtheta_1)_\zeta$, $(\ovtheta_{l + 1})_\zeta$, 
are shifted prefundamental representations $L^-_{1, \, x}$, $L^+_{l, \, x}$ 
and $L^+_{1, \, x}$, $L^-_{l, \, x}$ for certain values of the parameter~$x$, 
respectively, and all other prefundamental representations 
$L^\pm_{i, \, x}$, $i = 2, \ldots,l-1$, are presented in tensor products 
whose subquotients give the corresponding $q$-oscillator representations. 
Altogether, for a general higher rank case with $l \ge 2$, there are $2(l + 1)$ 
$q$-oscillator representations versus $2l$ prefundamental representations. 
The case of $l = 1$ is special. Here one has only $2$ representations of 
both kinds \cite{BooGoeKluNirRaz16}.

To be specific, let us gather the highest $\ell$-weights obtained in sections 
\ref{ss:a1}, \ref{ss:gen-a} and \ref{ss:l+1}. Thus, for the representation 
$(\theta_1)_\zeta$ we have
\begin{align*}
& \lambda_{\bm 0, \, 1} = -(l + 1) \, \omega_1, \\
& \Psi^+_{i, \, {\bf0}, \, 1}(u) = \left\{
\begin{array}{cl}
q^{- l - 1} \left( 1 - q^{- l} \, \zeta^s \, u \right)^{-1}, & i = 1, \\
1, & i = 2, \ldots, l,
\end{array}
\right. \\
\intertext{for the representations $(\theta_a)_\zeta$ with $a = 2, \ldots, l$ we have}
& \lambda_{\bm 0, \, a} = (l - a + 1) \, \omega_{a - 1} - (l - a + 2) \, \omega_a, \\ 
& \Psi^+_{i, \, {\bf0}, \, a}(u) = \left\{ 
\begin{array}{cl} 
1, & i = 1,\ldots,a-2, \\
q^{l - a + 1} \, \left( 1 - q^{- l + a} \, \zeta^s \, u \right), & i = a-1, \\
q^{- l + a - 2} \, \left( 1 - q^{- l + a - 1} \, \zeta^s \, u \right)^{-1}, & i = a, \\
1, & i = a + 1, \ldots, l,
\end{array}
\right. \\
\intertext{and for the representation $(\theta_{l + 1})_\zeta$ we have}
& \lambda_{\bm 0, \, l + 1} = 0, \\
& \Psi^+_{i, \, {\bf0}, \, l + 1}(u) = \left\{
\begin{array}{cl}
1, & i = 1, \ldots, l - 1, \\
1 - q \, \zeta^s \, u, & i = l.
\end{array}
\right.
\end{align*}
It means that
\begin{align*}
&(\theta_1)_\zeta \cong L^\ms_{\xi_1} \otimes L^-_{1, \, q^{- l} \zeta^s}\, , \\
&(\theta_a)_\zeta \cong L^\ms_{\xi_a} \otimes (L^+_{a - 1, \, q^{- l + a} \zeta^s} 
\ovotimes L^-_{a, \, q^{- l + a - 1} \zeta^s }), \qquad a = 2, \ldots, l, \\
&(\theta_{l + 1})_\zeta \cong L^+_{l, \, q \, \zeta^s}.
\end{align*}
For the definition of the operation $\ovotimes$ see section \ref{s:gibs}. The shifts 
$\xi_{a}$ have the form 
\begin{equation*}
\xi_1 = - (l + 1) \, \omega_1, \quad \xi_a = (l - a + 1) \, \omega_{a - 1} - (l - a + 2) \, \omega_a, 
\quad a = 2, \ldots, l, \quad \xi_{l + 1} = 0.
\end{equation*}

Actually, in view of the above relations, we can also write a reversed 
relation expressing the prefundamental representations as subquotients 
of tensor products of $q$-oscillator representations. Explicitly we have
\begin{equation*}
L_{\xi^-_i} \otimes L^-_{i, \, \zeta^s} \cong (\theta_1)_{q^{l + i - 1} \zeta^s} 
\ovotimes (\theta_2)_{q^{l + i - 3} \zeta^s} \ovotimes \ldots \ovotimes (\theta_i)_{q^{l - i + 1} \zeta^s},
\end{equation*}
where the elements $\xi^-_i$ are defined as
\begin{equation*}
\xi^-_i = - 2 \sum_{j = 1}^{i - 1} \omega_j - (l - i + 2) \, \omega_i,
\end{equation*}
and
\begin{equation*}
L_{\xi^+_i} \otimes L^+_{i, \, \zeta^s} \cong (\theta_{i + 1})_{q^{l - i - 1} \zeta^s} 
\ovotimes (\theta_{i + 2})_{q^{l - i - 3} \zeta^s} \ovotimes \ldots \ovotimes 
(\theta_{l + 1})_{q^{- l + i - 1} \zeta^s},
\end{equation*}
with the elements $\xi^+_i$ given by the equation
\begin{equation*}
\xi^+_i = (l - i) \, \omega_i - 2 \sum_{j = i + 1}^l \omega_j.
\end{equation*}
Similar relations can also be written for the representations $\ovtheta_a$. We see that 
any $\uqbp$-module in the category $\calO$ can be presented as a shifted subquotient 
of a tensor product of $q$-oscillator representations. Besides, we note that the 
corresponding highest $\ell$-weights are as simple as those of the prefundamental 
representations. And in this sense, the $q$-oscillator representations are no less 
fundamental than the prefundamental ones. 

We denote the $\uqbp$-modules corresponding to the representations $\theta_a$ defined 
in (\ref{thetaa}) by $W_a$, $a = 1, \ldots, l+1$, and consider the $\uqbp$-module 
$(W_1)_{\zeta_1} \otimes \ldots \otimes (W_{l+1})_{\zeta_{l+1}}$. 
Then, the tensor product of the highest $\ell$-weight vectors is an 
$\ell$-weight vector of $\ell$-weight determined by the functions
\begin{equation*}
\Psi^+_i(u) = q^{-2} \, \frac{1 - q^{- l + i + 1} \, \zeta^s_{i+1} \, u}
{1 - q^{- l + i - 1} \, \zeta^s_{i} \, u}, \qquad i = 1, \ldots, l.
\end{equation*}

Let now $\widetilde V^\lambda$ be the representation of the whole quantum loop algebra 
$\uqlsllpo$ constructed with the help of Jimbo's homomorphism \cite{NirRaz16b}. 
Consider its restriction to the Borel subalgebra $\uqbp$. We denote 
this restriction again by $\widetilde V^\lambda$. It can be 
shown\footnote{Work in progress, to appear elsewhere.} that 
the highest $\ell$-weight of the $\uqbp$-module 
$(\widetilde V^\lambda)_\zeta$ is determined by 
the functions
\begin{equation*}
\Psi^+_i(u) = q^{\lambda_i - \lambda_{i+1}} \, 
\frac{1 - q^{2\lambda_{i+1} - i + 1} \, \zeta^s \, u}
{1 - q^{2\lambda_{i} - i + 1} \, \zeta^s \, u}, \qquad 
i = 1, \ldots, l.
\end{equation*}
We see that if
\begin{equation*}
\zeta_i = q^{2(\lambda + \rho)_i / s} \, \zeta, \qquad \rho_i = \frac{l}{2} - i + 1,
\end{equation*}
where $\rho$ denotes here the half-sum of all positive roots,
\begin{equation*}
\rho = \big( \, l/2, \, l/2 - 1, \, \ldots, - l/2 \, \big),
\end{equation*}
then the submodule of $(W_1)_{\zeta_1} \otimes \ldots \otimes (W_{l+1})_{\zeta_{l+1}}$ 
generated by the tensor product of the highest $\ell$-weight vectors of $(W_a)_{\zeta_a}$,
$a = 1, \ldots, l + 1$, is isomorphic to the shifted module 
$(\widetilde V^\lambda)_\zeta[\xi]$, where the shift $\xi$ 
is 
\begin{equation*}
\xi = \sum_{i = 1}^l (\lambda_{i + 1} - \lambda_i - 2) \, \omega_i.
\end{equation*} 

A similar conclusion holds for $\ovtheta_a$ as well. Altogether, this observation
about the connection between the highest $\ell$-weights is useful for understanding
the structure of functional relations between the integrability objects.

As we have noted already, the representations $(\theta_a)_\zeta$ and $(\ovtheta_a)_\zeta$ 
are used to construct $Q$-operators which are necessary tools for the theory of quantum integrable systems. Correspondingly, we have two sets of $Q$-operators, $Q_a(\zeta)$ and $\ovQ_a(\zeta)$, with $a = 1, \ldots, l + 1$. Each $Q$-operator is defined as a twisted 
trace over the representation space of $(\theta_a)_\zeta$ or $(\ovtheta_a)_\zeta$, which 
is in fact the representation space of $l$ copies of the $q$-oscillators. In papers \cite{BazLukMenSta10, BazFraLukMenSta11}, for quantum integrable systems related to 
the Yangian $\mathrm Y(\gllpo)$, the complex of $Q$-operators $Q_I(z)$, where $I$ is 
an increasing set of integers $\{a_1, \, a_2, \, \ldots, \, a_p\} \subset \{1, \, 2, \, 
\ldots, \, l + 1\}$, was introduced. It is clear that here one has $2^{l + 1}$ $Q$-operators. The operator $Q_{\{a_1, \, a_2, \, \ldots, \, a_p\}}(z)$ is constructed as a twisted trace 
over the representation space of $p(l + 1 - p)$ copies of the harmonic oscillators. The operators $Q_{\{a\}}(z)$ and $Q_{\{1, \, 2, \, \ldots, \, l + 1\} \smallsetminus \{a\}}(z)$ 
are the limit for $q \to 1$ of the operators $Q_a(\zeta)$ and $\ovQ_a(\zeta)$, respectively. Although, by using the appropriate functional relations, any operator $Q_{\{a_1, \, a_2, \, \ldots, \, a_p\}}(z)$ can be expressed via the operators $Q_{\{a\}}(z)$, $a = 1, 2, \ldots, 
l + 1$, it would be interesting to find deformed analogues for all operators~$Q_{\{a_1, \, 
a_2, \, \ldots, \, a_p\}}(z)$.

\vspace{1em}

\noindent {\bf Acknowledgements.} This work was supported in part by the Deutsche Forschungsgemeinschaft in the framework of the research group FOR 2316, by the DFG 
grant KL \hbox{645/10-1}, and by the RFBR grants \#~14-01-91335 and \#~16-01-00473.

\newcommand{\noopsort}[1]{}
\providecommand{\bysame}{\leavevmode\hbox to3em{\hrulefill}\thinspace}
\providecommand{\href}[2]{#2}
\providecommand{\curlanguage}[1]{%
 \expandafter\ifx\csname #1\endcsname\relax
 \else\csname #1\endcsname\fi}

\end{document}